\documentclass[aps,prl,reprint,nofootinbib,numbers,nobalancelastpage]{revtex4-1}
\usepackage{graphicx}
\usepackage{blindtext}
\usepackage{braket}
\usepackage{mathtools}
\usepackage{amsmath}
\usepackage{amssymb}
\usepackage{cancel}
\usepackage{empheq}
\usepackage{bbm}
\usepackage[strict]{changepage}
\usepackage{xcolor,soul}
\sethlcolor{yellow}
\usepackage{natbib}
\usepackage[scr]{rsfso}
\usepackage[colorlinks=true]{hyperref}
\usepackage[sort&compress]{natbib}
\usepackage{etoolbox}

\begin{document}
\title{A Comprehensive Analytical 
Model of the Dynamic Z-Pinch}
\author{Alejandro Mesa Dame$^{\dagger}$$^{1,2}$, Eric S. Lavine$^{1}$, David A. Hammer$^{1}$}
\affiliation{$^{1}$Laboratory of Plasma Studies, Cornell University, Ithaca, NY 14850\\
$^{2}$Princeton Plasma Physics Laboratory, Princeton University, Princeton, NJ 08540}
\footnotetext{$^{\dagger}$\,Corresponding author: amesadam@pppl.gov}
\begin{abstract}

We present an analytical 1D axisymmetric model describing the evolution of the dynamic z-pinch. This model is capable of predicting the trajectories of the imploding sheath's magnetic piston and preceding shock front, along with the velocity, pressure, density, and magnetic field profiles, for any time-dependent current, spatially varying initial density profile, and weak initial axial field. The implosion is divided into stages, with each stage described by a set of coupled ordinary differential equations derived from the ideal MHD equations. Comparisons with experimental data from the COBRA pulsed-power facility are quite promising and imply this model could prove useful in designing and analyzing future pulsed-power experiments.
\end{abstract}

\maketitle
\section{I. Introduction}
\label{sec:intro}
\vspace{-10pt}
The z-pinch~\cite{Haines1982,Liberman1999,Haines2000,Ryutov2000,Haines2011} is a classic plasma confinement configuration with axial symmetry. It consists of running an axial current through a cylindrical load, sometimes in the presence of an applied axial magnetic field. This current forms into a thin cylindrical sheath that interacts with its own induced azimuthal magnetic field to produce a radially inward pinch force. The resulting runaway effect causes the column to violently collapse, ultimately heating and compressing the fuel inside into a high energy density plasma. Z-pinches are excellent high-intensity x-ray~\cite{Leeper1999,Matzen1997,Matzen1999,Spielman2000,Haines2005,Giuliani2015} and neutron sources~\cite{Coverdale2007,Giuliani2015,Zhang2019}, and are used for a variety of applications including the study of fundamental plasma physics and research on controlled fusion via inertial (ICF)~\cite{Leeper1999,Matzen1999,Spielman2000,Haines2005} and magneto-inertial (MIF)~\cite{Shumlak2020,Yager2022} confinement. \\
\indent At the Cornell Laboratory of Plasma Studies, the COBRA~\cite{Ouart2016,Qi2018,Lavine2021,Lavine2022,Lavine2024,Lavine2025} pulsed-power generator is used together with computer simulation tools and advanced diagnostics to study the fundamental properties and underlying physics of high energy density plasmas (HEDPs). One type of z-pinch conducted on COBRA is the gas-puff z-pinch~\cite{Shiloh1978,Felber1988,Coverdale2007,Giuliani2015,Shah2024} where the load consists of two concentric annular puffs and a central jet of neon, argon, or krypton gas. The initial density profile is controlled by adjusting the relative plenum pressures of the puffs~\cite{PLIF}. The load is preionized to prevent power reflections and unacceptably large voltages early in the discharge~\cite{Lavine2025}. \\
\indent Shot time on COBRA is limited and requires support staff, contributing to operational costs. To optimize experimental campaigns, it is crucial that researchers can anticipate when and where key physical phenomena occur. This enables precise targeting and timing of experimental diagnostics such as Thomson scattering~\cite{Glenzer2009,Rocco2018,Rocco2022,Zorondo2022,Banasek2023}, laser interferometry~\cite{Qi2002,Chen2009}, and self-emission imaging. To support these efforts, models must be developed that can rapidly and accurately predict implosion dynamics for a variety of operation scenarios.\\
\indent Currently, such predictions are made with a combination of past data and 2D axisymmetric magnetohydrodynamic (MHD) simulation codes such as GORGON~\cite{Chittenden2001,Chittenden2004,Ciardi2007,Chittenden2008,Bocchi2013}, developed at Imperial College, and PERSEUS~\cite{Martin2010,Seyler2011,Hamlin,Woolstrum2020,Woolstrum2022}, developed at Cornell. While such models are invaluable for interpreting experimental results, they are computationally intensive, requiring hours of runtime on dedicated machines, even for simplified scenarios. Moreover, they demand detailed input data regarding initial and boundary conditions, which is not always available. This makes MHD codes impractical for guiding decisions during an ongoing shot campaign or for getting quick estimates of timings and physical parameters.  Considering that the ultimate purpose of pulsed-power research is an improved understanding of fundamental plasma physics, producing numerical solutions with a general purpose MHD code is not always helpful for developing intuition and deciding what avenues of research to pursue next.\\
\indent The aim of this research was to develop a faster, simpler code that employs a 1D analytical model of the dynamic z-pinch, derived from first principles, which reflects and predicts experimental results with reasonable accuracy. This model is able to predict the trajectories of the imploding sheath’s inner ``shock” and outer ``piston” radii, along with the velocity, pressure, density, and magnetic field profiles everywhere, for any initial density profile, current waveform, and weak applied axial field. The implosion is divided into stages, with each stage described by a pair of coupled ordinary differential equations derived from the ideal MHD equations. These equations are solved via numerical integration in one dimension.\\
\indent Analytical models of the dynamic z-pinch have been previously examined, starting in the early 1950s with Rosenbluth's~\cite{Rosenbluth,Miyamoto} snowplow model, which predicts the implosion trajectory of a perfectly conducting liner described by a single time-dependent radial coordinate. Potter's~\cite{Potter} strong-shock slug model extended the snowplow model, allowing the liner to have a finite thickness between two radii with coupled movements, the piston and shock, and described the structure between them. Piriz et al.~\cite{Piriz} presented an alternative to the slug and snowplow models derived directly from the ideal MHD equations. Angus et al.~\cite{Angus} corrected mathematical errors in the Piriz model and generalized it for spatially varying initial density profile and time-dependent current. However, the Angus model has a singular point at the rigorous initial conditions in Eq.~(\ref{Eq:IC}) where the shock and piston radii are equal, and thus requires ad-hoc initialization. It also lacks a complete analytical description of the fluid element motions as well as the velocity, pressure, density, and magnetic field profiles, and ceases to apply beyond the moment of incidence, when the shock reaches the axis. Furthermore, no physically self-consistent analytical model to date has considered implosion in the simultaneous presence of a spatially varying initial density profile, time-dependent current, and applied axial magnetic field as is often the case for real z-pinch experiments.\\
\indent We therefore generalize the derivation in Angus et al.~\cite{Angus}, presenting explicit expressions for the governing equations for any initial density profile, current waveform, and weak applied axial field. We propose an ordering of implosion stages, taking inspiration from Lee and Saw's multi-stage Lee model for the dense plasma focus (DPF)~\cite{Lee}, designed to avoid the divergent initial conditions of the Angus model, extend the model beyond the moment of incidence, and best reflect the underlying physics at each stage of the implosion. We then compare this model to experimental results from the COBRA pulsed-power generator and demonstrate close agreement with the data. While Angus compared predicted radial trajectories with an MHD code, no comparison of this kind has been made between any analytical model beyond the basic snowplow~\cite{Giuliani2015,Lavine2021,Lavine2022,Lavine2024} and experimental data from pulsed-power facilities. Even comparisons with MHD codes were done only in the case of constant density and current~\cite{Angus}.\\
\indent The outline of the paper is as follows. We begin in Sec.\,\hyperref[sec:config]{II} by illustrating the configuration of the dynamic z-pinch in full detail. In Sec.\,\hyperref[sec:stages]{III}, we describe the implosion of the z-pinch as a series of stages, identify the events that separate these stages, and finally state the models that characterize them. In Sec.\,\hyperref[sec:model]{IV}, we present each of these models, explain their physical assumptions, and explicitly give their governing ODEs. In Sec.\,\hyperref[sec:comparison]{V}, we fit the adiabatic index and initial radius of the liner in our model to a single COBRA shot. Then, we compare this fitted model to data from several other shots with different initial density profiles, current waveforms, and axial fields to evaluate its effectiveness. In Sec.\,\hyperref[sec:profiles]{VI}, we obtain predicted velocity, pressure, density, and axial field profiles within the sheath and compute results for several shots. In Sec.\,\hyperref[sec:fundamental]{VII}, we revisit the fundamental relation governing the z-pinch to understand the core principles at play. In Sec.\,\hyperref[sec:conclusions]{VIII}, we present our conclusions.
\begin{figure}
\centering
\noindent\includegraphics[width=0.8\columnwidth,height=0.8\columnwidth]{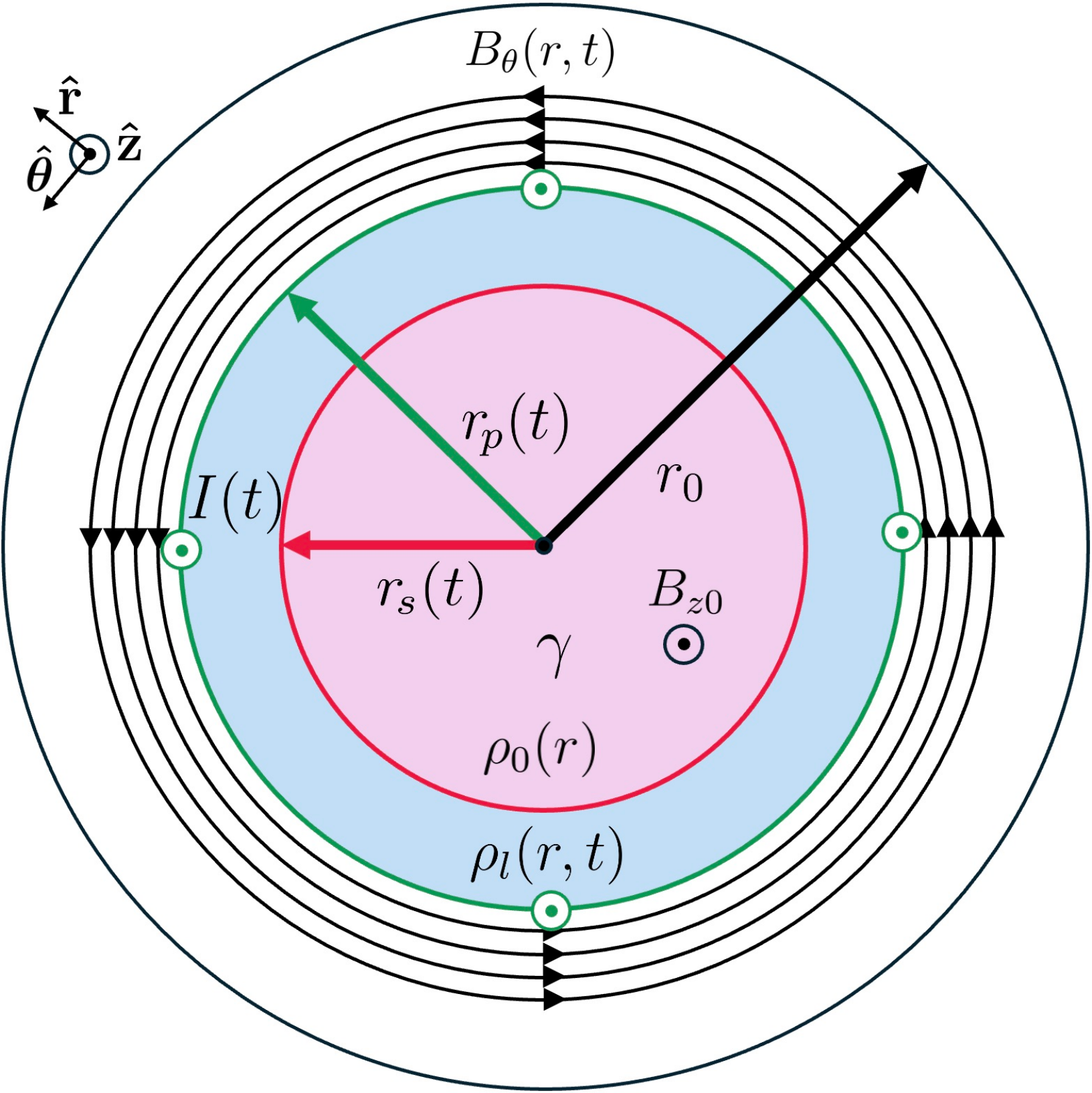}
\caption{Z-Pinch Configuration. The z-pinch is shown in cylindrical geometry with time-dependent current $I(t)$, pinching azimuthal field $B_{\theta}(r,t)$, spatially varying initial density $\rho_{0}(r)$, initial axial field $B_{z0}$, and adiabatic index $\gamma$. The piston $r_{p}(t)$ and shock $r_{s}(t)$ radii separate the system into three distinct radial regions, with the sheath density $\rho_{l}(r,t)$ between them.\label{Fig:config}}
\end{figure}
\vspace{-10pt}
\section{II. Configuration}
\vspace{-5pt}
\label{sec:config}
The configuration for the dynamic z-pinch is illustrated in Fig.~\ref{Fig:config} in standard cylindrical coordinates ($r,\theta,z$). We assume that our system is axisymmetric and translationally symmetric such that $\partial/\partial\theta = \partial/\partial z = 0$, reducing our problem to one dimension.\\
\indent Within a hollow cylinder of radius $r_{0}$, preionized gas with initial radial mass distribution $\rho_{0}(r)$ is pumped into the gap between two vertically separated electrodes and magnetized with uniform axial field $B_{z0}$. This gas serves as the medium to conduct an externally applied axial current profile $I(t)$ in the positive $\mathbf{\hat{z}}$ direction. The resulting current density $\vec{J}$ resides almost entirely at the outer surface and produces an azimuthal magnetic field $B_{\theta}(r,t)$ in the positive $\boldsymbol{\hat{\theta}}$ direction around itself via Amp\`ere's law. Together, the current and magnetic field produce a $\vec{J}\times\vec{B}$ force in the negative $\mathbf{\hat{r}}$ direction, leading to a constriction of the material, which is known as the pinch force. (In the presence of an axial field, an azimuthal current flows as well, generating an opposing force.) The outermost radius moving inwards under this force, where the axial current density overwhelmingly resides, is referred to as the ``piston" and picks up the vast majority of the mass it encounters on its inward journey like a snowplow. The preceding shock wave the piston produces is referred to as the ``shock". Across the shock front, quantities of interest suffer characteristic discontinuities that can be described via the Rankine-Hugoniot (RH) jump conditions~\cite{Classic,Zeldovich}, as illustrated in Appendix~\hyperref[App:A]{A}. The region between the piston and shock is referred to interchangeably as the ``liner" or ``sheath".
\vspace{-10pt}
\section{III. Implosion Stages}
\begin{figure}
\centering
\noindent\includegraphics[width=\columnwidth]{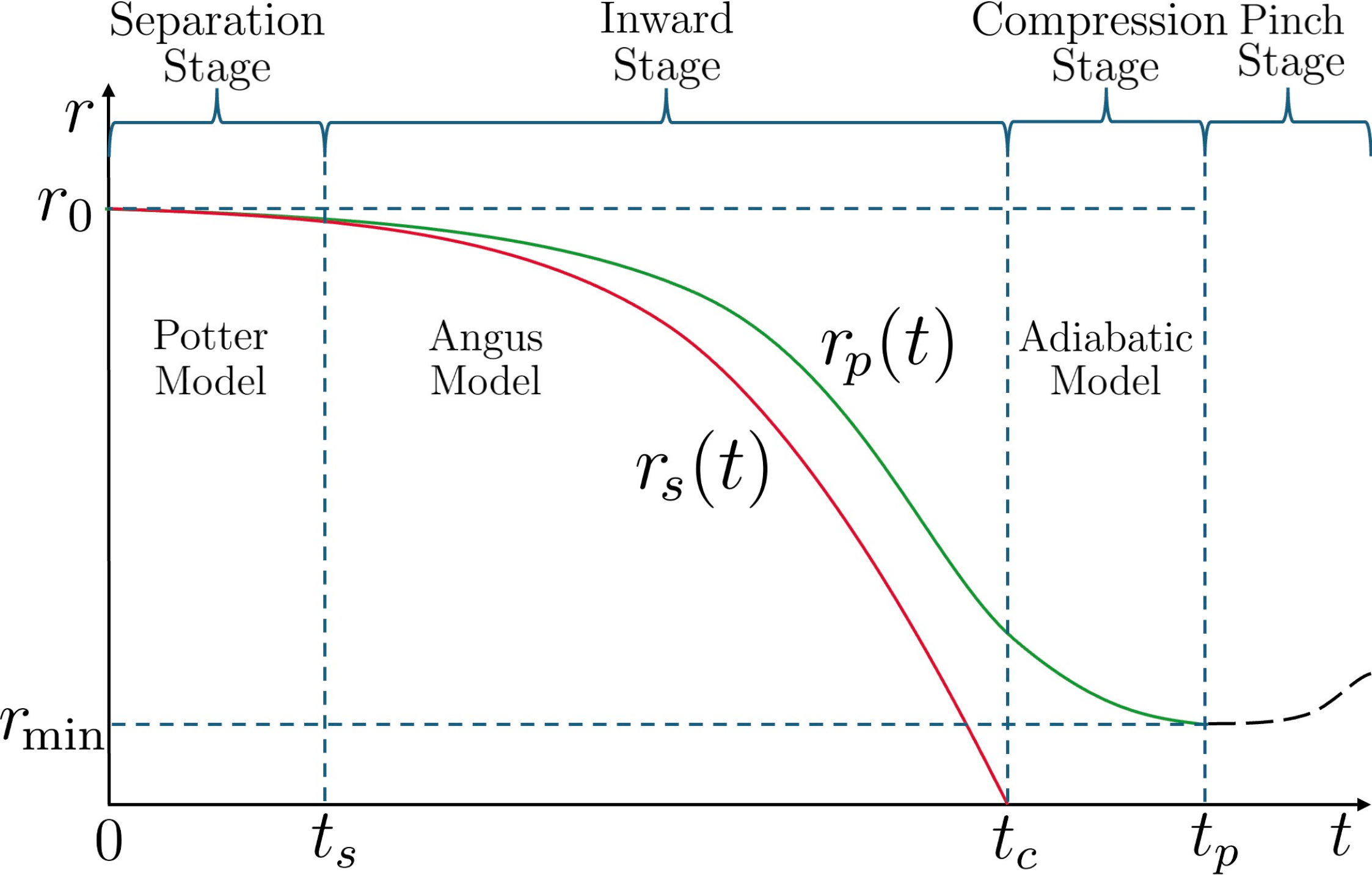}
\caption{Implosion Stages. Here we see the evolution of the piston and shock radii throughout various stages of the implosion. Each stage is associated with a characteristic model which best describes its underlying physics. Different models either asymptotically agree or are matched with physical boundary conditions at transition times.\label{Fig:stages}}
\end{figure}

\vspace{-5pt}
\label{sec:stages}
\indent Before we begin modeling its evolution, let us first describe and give names to the various stages of the implosion as shown in Fig.~\ref{Fig:stages}.\\
\indent At $t = 0$, the current waveform $I(t)$ rises from zero as the piston and shock simultaneously form at $r = r_{0}$, initiating the separation stage. They remain close together as the implosion commences, bounding a thin sheath with negligible inertia. The piston starts moving inwards under the $\vec{J}\times\vec{B}$ force, which approximately equilibrates with the material pressure, as the sheath expands adiabatically. Potter's slug model is the most comprehensive model which is well-defined for $r_{s} = r_{p}$ but maintains a distinction  between the piston and shock. It is also the limit of the Angus model for early times as shown in Appendix~\hyperref[F:IV]{F.IV}, so we use it here to model this separation stage, as illustrated in Appendix~\hyperref[App:E]{E}.\\
\indent At the separation time $t = t_{s}$, the piston and shock become fully distinguishable, and the inertia of the sheath becomes appreciable, signaling a transition to the inward stage. The piston accelerates inwards under the $\vec{J}\times\vec{B}$ force and accumulates mass, while the shock front moves inwards ahead of it. The inward motions of the piston and shock are most comprehensively described by the Angus model, as illustrated in Appendix~\hyperref[App:F]{F}.\\
\indent At the incidence time $t = t_{c}$, the shock reaches the central axis, marking the beginning of the compression stage. In the limit of infinite sound speed in the sheath, there is no reflected shock and the on-axis boundary conditions are communicated to the entire column instantly. The piston continues to accelerate inwards under the $\vec{J}\times\vec{B}$ force and the whole column enclosed by the piston is thus compressed adiabatically. This is sufficient to determine the motion of the piston, as illustrated in Appendix~\hyperref[App:G]{G}.\\
\indent At the stagnation time $t = t_{p}$, the material pressure finally overpowers the $\vec{J}\times\vec{B}$ force and inertia of the sheath, and the piston comes to rest at $r_{\text{min}} \equiv r_{p}(t_{p})$, thereby initiating the pinch stage. Energy loss, due to the now significant production of x-rays and neutrons, makes this stage much too complex to model analytically~\cite{Narkis2019,Lee}.
\vspace{-5pt}
\section{IV. The Model}
\label{sec:model}
\indent The standard ideal MHD system is given as follows,
\begin{align}\label{Eq:MHD-1}&\frac{\partial\rho}{\partial t}+\vec{\nabla}\cdot(\rho\vec{v})=0,\\
\label{Eq:MHD-2}&\rho\!\left(\!\frac{\partial}{\partial t}\!+\!\vec{v}\cdot\vec{\nabla}\!\right)\!\vec{v}=\frac{(\vec{\nabla}\times\vec{B})\times\vec{B}}{\mu_{0}}\!-\!\vec{\nabla}P,\\
\label{Eq:MHD-3}&\left(\!\frac{\partial}{\partial t}\!+\!\vec{v}\cdot\vec{\nabla}\!\right)\!P=-\gamma P\vec{\nabla}\cdot\vec{v},\\
\label{Eq:MHD-4}&\frac{\partial\vec{B}}{\partial t} = \vec{\nabla}\times(\vec{v}\times\vec{B}).\end{align}
In each stage, we derive a pair of coupled ODEs in the piston radius, $r_{p}$, and the shock radius, $r_{s}$, from Eqs.~(\ref{Eq:MHD-2},\ref{Eq:MHD-3}). We simultaneously impose conservation of mass from Eq.~(\ref{Eq:MHD-1}) and a frozen-in axial field according to Alfv\'en's theorem from Eq.~(\ref{Eq:MHD-4}).\\
\indent The general principles that govern our model can be summarized as follows.
The Mach number of the inward shock relative to the initial, untouched, upstream region's sound and Alfv\'en speeds should be large, and the strong-shock limit of the RH jump conditions applies. This ordering is discussed in Appendix~\hyperref[App:A]{A}, where it is determined that the axial field does not significantly affect the RH jump conditions.\\
\indent Furthermore, since the sound speed is high in the postshock plasma, it stands to reason that any significant pressure gradients are quickly resolved. We therefore make the simplifying assumption of a uniform pressure profile within this region. The adiabatic law in Eq.~(\ref{Eq:MHD-3}) and the RH jump conditions from Appendix~\hyperref[App:A]{A} are then sufficient to determine the velocity and pressure profiles as shown in Appendices~\hyperref[App:B]{B} and \hyperref[App:C]{C}, respectively. These profiles suggest that, in the time leading up to the moment of incidence, all of the incoming shock's energy is carried away by sound waves, precluding the need to consider a reflected shock, as shown in Appendix~\hyperref[B:IV]{B.IV}.

Since the shock is the only entropic agent in our system, individual fluid elements are expected to evolve adiabatically on either side of the shock front. Conservation of mass from Eq.~(\ref{Eq:MHD-1}), together with the pressure profiles, is sufficient to determine their densities and motions, and Alfv\'en's theorem in Eq.~(\ref{Eq:MHD-4}) consequently determines their axial fields, as shown in Appendix~\hyperref[App:D]{D}. The infinite density which naturally arises at the piston under these simplifying assumptions is abused to justify the pressure discontinuity required for inward acceleration despite the uniform pressure profile within the sheath itself.\\
\indent Let us use subscripts $0$ and $l$ to denote the initial and postshock/liner regions, respectively. Then the velocity, pressure, density, and magnetic field profiles throughout the implosion are given as follows,
\begin{align}\label{Eq:model1}&\vec{v}(r,t) = \begin{cases}0,&r_{p}(t)<r\leq r_{0}\\v_{l}(r,t)\mathbf{\hat{r}},&r_{s}(t)\leq r \leq r_{p}(t)\\0,&0\leq r < r_{s}(t)\end{cases},\end{align}
\begin{align}\label{Eq:model2}&P(r,t) = \begin{cases}0,&r_{p}(t)<r\leq r_{0}\\P_{l}(t),&r_{s}(t)\leq r \leq r_{p}(t)\\P_{0}(r),&0\leq r < r_{s}(t)\end{cases},\end{align}
\begin{align}\label{Eq:model3}&\rho(r,t) = \begin{cases}0,&r_{p}(t)<r\leq r_{0}\\\rho_{l}(r,t),&r_{s}(t)\leq r \leq r_{p}(t)\\\rho_{0}(r),&0\leq r < r_{s}(t)\end{cases},\end{align}
\begin{align}\label{Eq:model4}&\Vec{B}(r,t)=\begin{cases}B_{\theta}(r,t)\boldsymbol{\hat{\theta}}& r_{p}(t) < r \leq r_{0}\\B_{zl}(r,t)\mathbf{\hat{z}} & r_{s}(t)\leq r\leq r_{p}(t)\\B_{z0}\mathbf{\hat{z}} & 0 \leq r < r_{s}(t) \end{cases},\end{align}
where we can define $r_{s}(t \geq t_{c}) \equiv 0$ to extend the validity of Eqs.~(\ref{Eq:model1}-\ref{Eq:model4}) to all times.
One can verify that, as must be the case, $\vec{v}(r,0) \!=\! 0,\, P(r,0) \!=\! P_{0}(r),\,\rho(r,0) \!=\! \rho_{0}(r)$, and $\vec{B}(r,0) \!=\! B_{z0}\mathbf{\hat{z}}$. The piston and shock radii are subject to initial conditions,
\begin{align}\label{Eq:IC}&r_{p}(0) = r_{s}(0) = r_{0},\quad\dot{r}_{p}(0) = \dot{r}_{s}(0) = 0.\end{align}

\subsection*{i. Separation Stage $(0 \leq t < t_{s})$}
Generalizing Potter's model as shown in Appendix~\hyperref[App:E]{E}, we obtain the coupled set of first-order ODEs,
\begin{align}\label{Eq:Potter}&\dot{r}_{s}\!=\!-\frac{I}{4\pi r_{p}\!}\sqrt{\!\frac{\mu_{0}(\gamma\!+\!1)}{\rho_{0}(r_{s})}},\,\dot{r}_{p}=\frac{\frac{2}{\gamma+1}r_{s}\dot{r}_{s}-\frac{r_{p}^{2}-r_{s}^{2}}{\gamma}\frac{\dot{I}}{I}}{r_{p}-\frac{r_{p}^{2}-r_{s}^{2}}{\gamma r_{p}}}.\!\!\end{align}
Eq.~(\ref{Eq:Potter}) is the generalization of Potter's~\cite{Potter} Eqs.~(11,12) for spatially varying initial density and time-dependent current. The sheath inertia and axial field present negligible contributions during this short initial stage since the piston has not yet accumulated enough mass for their effects to be palpable.
\subsection*{ii. Inward Stage $(t_{s} \leq t < t_{c})$}
Generalizing Angus' model as shown in Appendix~\hyperref[App:F]{F}, we arrive at the coupled set of second-order ODEs,
\begin{align}
\label{Eq:simple-rp}&\!\!\!\!\!\!\!\!\!\frac{m(t)}{2\pi}r_{p}\ddot{r}_{p}\!=\!-\frac{\mu_{0}I^{2}}{8\pi^{2}}\!+\!\frac{2}{\gamma\!+\!1}\rho_{0}(r_{s})\dot{r}_{s}^{2}r_{p}^{2}\!+\!2\!\!\int\limits_{r_{s}}^{r_{p}}\!\frac{B_{zl}^{2}(r,t)\!}{2\mu_{0}}rdr,\!\!\!\!\!\\
\label{Eq:Angus-rs}
&\ddot{r}_s \!= -\frac{\gamma\dot{r}_{s}}{r_{p}^{2}\!-\!r_{s}^{2}}\!\left(\!r_{p}\dot{r}_{p}\!-\!\frac{2}{\gamma\!+\!1}r_{s}\dot{r}_{s}\!\right)\!\!-\!\frac{\dot{r}_{s}}{2}\frac{d}{dt}\ln\rho_{0}(r_{s}),
\end{align}
where the sheath mass per unit length $m(t)$ is given by,
\begin{align}
\label{Eq:Angus-greek}&m(t) \equiv \!\int\limits_{0}^{2\pi}\!\int\limits_{r_{s}}^{r_{p}}\rho_{l}(r,t)rdr d\theta = 2\pi\!\!\int\limits_{r_{s}}^{r_{0}}\!\!\rho_{0}(r)rdr.
\end{align}
Eqs.~(\ref{Eq:simple-rp},\ref{Eq:Angus-rs}) are analogous to Angus et al.~\cite{Angus} Eqs.~(33,25). The axial field profile $B_{zl}(r,t)$ can be determined via Alfv\'en's theorem in Eq.~(\ref{Eq:MHD-4}) as shown in Appendix~\hyperref[App:D]{D}.
\subsection*{iii. Compression Stage $(t_{c} \!\leq\! t \!\leq\! t_{p})$}
Extending the Angus model and imposing adiabatic evolution of the column enclosed by the piston as shown in Appendix~\hyperref[App:G]{G}, we arrive at the following ODE,
\begin{align}\label{Eq:Adiabatic-rp}
&\frac{m(t)}{2\pi}r_{p}\ddot{r}_{p}\!=\!-\frac{\mu_{0}I^{2}}{8\pi^{2}}\!+\!\frac{2}{\gamma\!+\!1}\rho_{0}(0)\dot{r}_{s}(t_{c}^{-})^{2}\!\!\left(\!\frac{r_{p}(t_{c})}{r_{p}(t)}\!\right)^{\!\!2\gamma}\!\!\!\!\!r_{p}^{2}\nonumber\\
&\!+2\!\int\limits_{0}^{r_{p}}\!\frac{B_{zl}^{2}(r,t)}{2\mu_{0}}rdr.\end{align}
Eq.~(\ref{Eq:Adiabatic-rp}) is simply the continuation of Eq.~(\ref{Eq:simple-rp}) for the updated pressure profile during this stage.
\subsection*{iv. Conservation of Energy}
Generalizing the conservation of energy criterion in Angus et al.~\cite{Angus} Eq.~(17) as shown in Appendix~\hyperref[F:II]{F.II},
\begin{align}\label{Eq:Angus-Wm}&\int\limits_{r_{s}}^{r_{p}}\!\left(\!\frac{2P(r,t)}{\gamma\!-\!1}\!+\!\frac{B_{z}^{2}(r,t)}{2\mu_{0}}\!\right)rdr\simeq\!\!\int\limits_{r_{p}}^{r_{0}}\!\frac{B_{\theta}^{2}(r_{p}^{+}\!,t)}{2\mu_{0}}r_{p}dr_{p}.\end{align}
This states that the sum of the total mechanical and magnetic energies of the sheath is approximately equal to the total work done by the azimuthal field. Once again, defining $r_{s}(t\geq t_{c}) \equiv 0$ allows us to extend its validity to all times.\\
\indent Two conjugate versions of Eqs.~(\ref{Eq:simple-rp},\ref{Eq:Adiabatic-rp}) obtained via Eq.~(\ref{Eq:Angus-Wm}) are given in Eqs.~(\ref{F:angus-rp},\ref{F:pre-novel}) and Eqs.~(\ref{G:angus-rp},\ref{G:pre-novel}). All three versions perform similarly, but each possesses distinct advantages from an implementation standpoint. Using Eqs.~(\ref{F:pre-novel},\ref{G:pre-novel}) for example does not require explicit knowledge of the axial field or tracking of the fluid elements. Eqs.~(\ref{Eq:simple-rp},\ref{Eq:Adiabatic-rp}) are the most easily interpretable as well as the most compatible with Eq.~(\ref{Eq:Potter}), as shown in Appendix~\hyperref[F:IV]{F.IV}.\\
\indent After $t = t_{p}$, this conservation of energy relation breaks down because the material pressure energy of the column starts being significantly converted into x-rays and neutrons. Observing the difference between the continuation of Eq.~(\ref{Eq:Adiabatic-rp}) for $t \geq t_{p}$ and experimental results may be useful for estimating the emitted energy.

\section{V. Experimental Comparisons}
\label{sec:comparison}
Now that we have presented our model, it is time to test it against experimental observations from argon gas-puff z-pinches at the COBRA pulsed-power installation.
\vspace{-20pt}
\subsection{i. Parameter Calibration}
\vspace{-5pt}
When comparing with real z-pinch shots, the known quantities are typically the initial density distribution $\rho_{0}(r)$, the current waveform $I(t)$, and the applied axial magnetic field $B_{z0}$, since these are externally imposed. However, the starting radius $r_{0}$ at which the liner forms and the adiabatic index $\gamma$ are unknown. As a first guess, $r_{0}$ can be taken to be $3.50$cm, the outer diameter of the gas nozzle, and $\gamma$ will be somewhat less than the idealized $5/3$ associated with three degrees of freedom due to ionization effects. However, incorrect choice of $(r_{0},\gamma)$ can lead to suboptimal predictions for such a sensitive and explosive nonlinear process. One way to determine these parameters is to use one shot to mean-squared-error (MSE) fit the pair $(r_{0},\!\gamma)$ based on limited experimental observations of the radial trajectories. Allowing for these two degrees of freedom also compensates for much of the natural error associated with using a simple analytical model and optimizes its performance for a particular device. After a single calibration, the model may be used freely. This calibration can be done as follows.\\
\indent Given $n_{p}$ experimental datapoints for the piston radius $X_{k} \equiv (t_{p,k},R_{p,k},\sigma_{p,k})$ and $n_{s}$ experimental datapoints for the shock radius $Y_{k} \equiv (t_{s,k},R_{s,k},\sigma_{s,k})$  of the form: (time, position, uncertainty), let us define the MSE loss function for simulated trajectories $r_{p}(t|r_{0},\!\gamma)$ and $r_{s}(t|r_{0},\!\gamma)$,
\begin{align}&L(\vec{X},\vec{Y},r_{0},\gamma) \!=\! \sum\limits_{\alpha = p,s}\sum\limits_{k=1}^{n_{\alpha}}\frac{(R_{\alpha,k}\!-\!r_{\alpha}(t_{\alpha,k}|r_{0},\!\gamma)\!)^{2}}{\sigma_{\alpha,k}^{2}}.\end{align}
We minimize this function with respect to combinations $(r_{0},\!\gamma)$ via direct calculation for values near $(3.50\text{cm},5/3)$.\\
\indent For the purposes of the following comparisons, the model was calibrated on COBRA shot 5532 using various diagnostics as shown in Figs.~\ref{Fig:5532-input} and \ref{Fig:5532-r} to obtain $(r_{0},\gamma) = (3.50\text{cm},1.37)$. These values appear quite reasonable since they respectively imply that the liner forms right at the outer edge of the supersonic gas puff nozzle and present a realistic adiabatic index for highly ionized argon gas~\cite{Hsu2018}. We also note that all of the employed initial density profiles are based on a single standardized planar laser-induced fluorescence (PLIF) measurement~\cite{PLIF}, which is then linearly scaled for each shot based on the plenum pressures of the three nozzles.
\vspace{-5pt}
\subsection{ii. Shot Comparison (Low vs High Initial Density)}
\begin{figure}[h!tbp]
\centering
\noindent\includegraphics[width=\columnwidth]{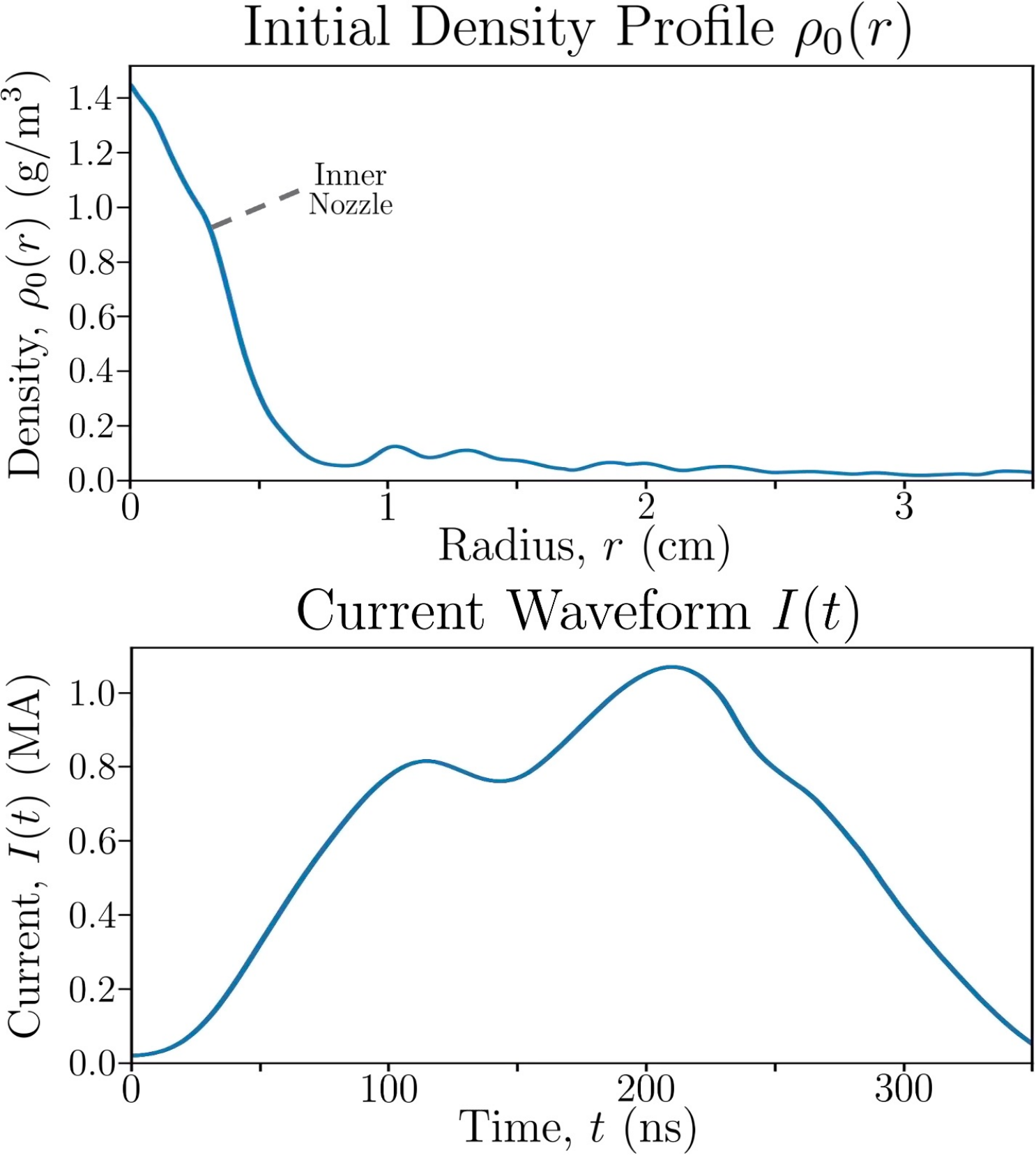}
\caption{Initial Density and Current Waveform for Shot 5532. Here we see the initial density profile $\rho_{0}(r)$ and current waveform $I(t)$ for shot 5532 as measured by planar laser-induced fluorescence (PLIF)~\cite{PLIF} and a Rogowski coil around the load, respectively.\label{Fig:5532-input}}
\end{figure}

\begin{figure}[h!tbp]
\centering
\noindent\includegraphics[width=\columnwidth]{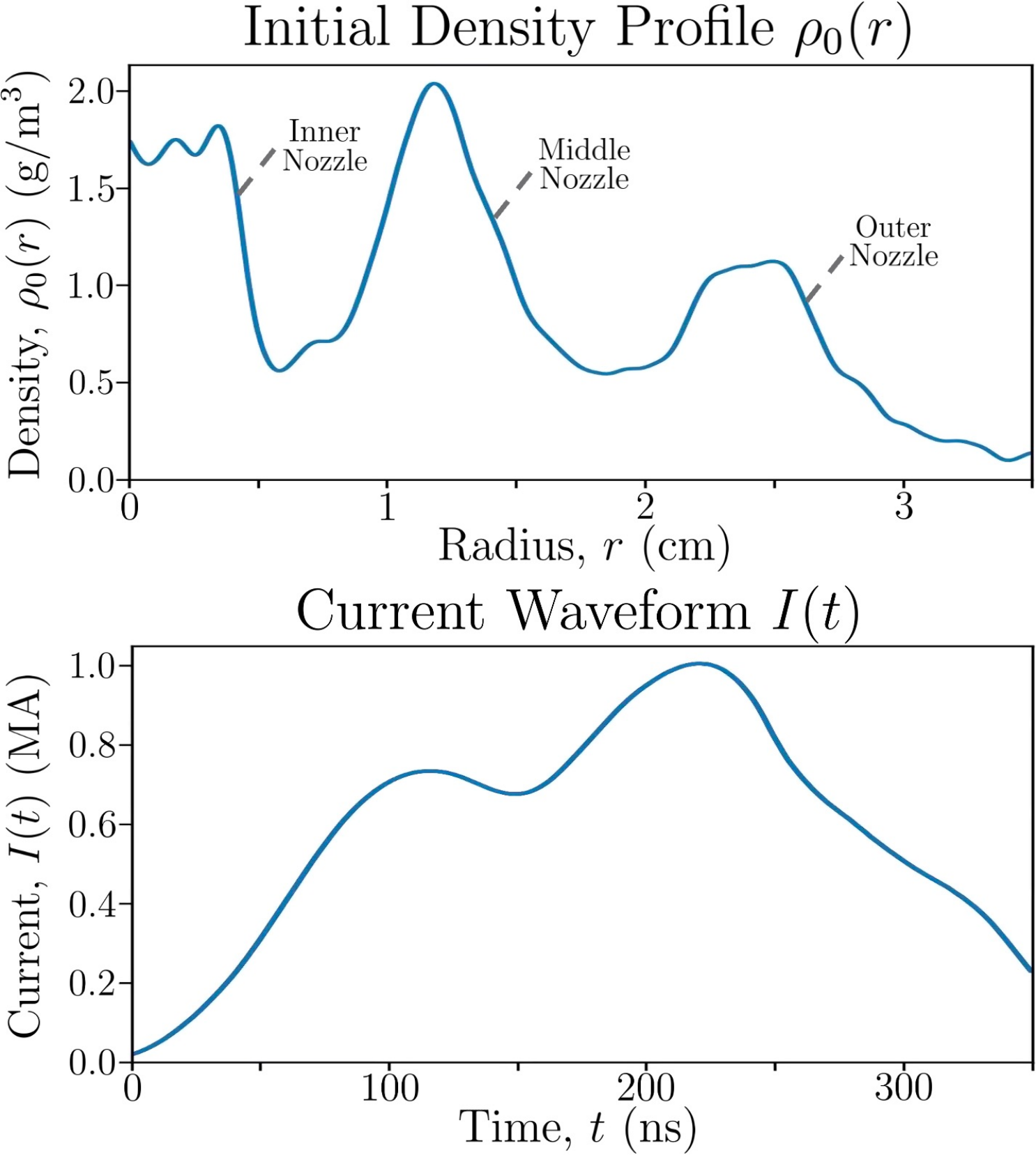}
\caption{Initial Density and Current Waveform for Shot 4959. Here we see the initial density profile $\rho_{0}(r)$ and current waveform $I(t)$ for shot 4959 as measured by planar laser-induced fluorescence (PLIF)~\cite{PLIF} and a Rogowski coil around the load, respectively.\label{Fig:4959-input}}
\end{figure}

\vspace{-5pt}
\indent As can be seen in Fig.~\ref{Fig:5532-input}, the density profile for the calibration shot 5532 is dominated by the central nozzle, with lower fill densities provided by the middle and outer nozzles. This shot was prepared such that the incidence time coincided with the peak axial current. We then tested the model on shot 4959, which employed a similar current waveform but featured a higher, more intricate initial density profile with significant contributions from all three nozzles, as shown in Fig.~\ref{Fig:4959-input}. The results are shown in Figs.~\ref{Fig:5532-r} and \ref{Fig:4959-r}. The model predictions agree closely with the data.
\vspace{-5pt}
\subsection{iii. Shot Comparison ($B_{z0} = 0$ vs $B_{z0} = 0.2\text{T}$)}
\begin{figure}[h!tbp]
\centering
\noindent\includegraphics[width=\columnwidth]{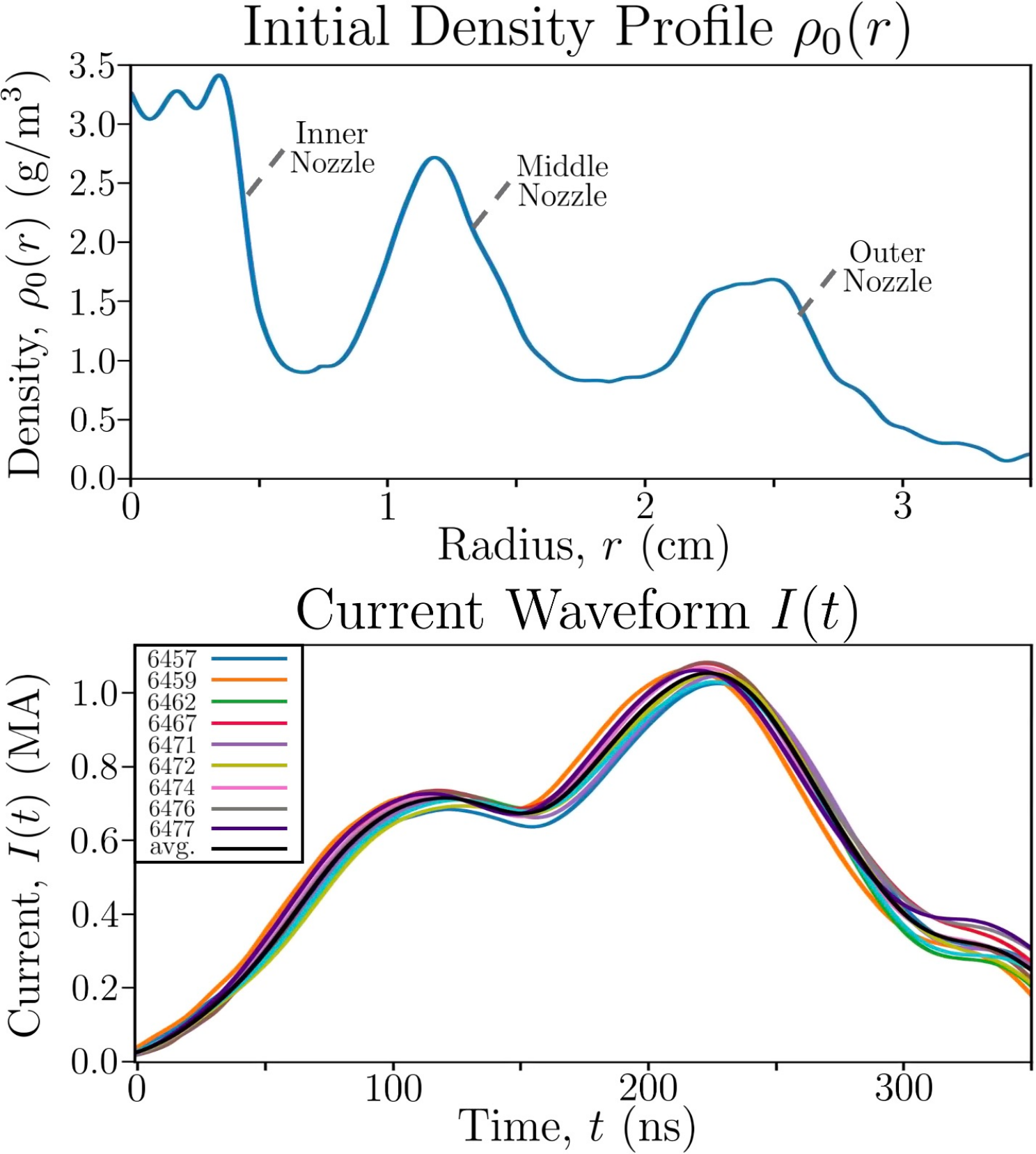}
\caption{Initial Density and Current Waveform for the $B_{z0}=0$ and $B_{z0}=0.2\text{T}$ Ensembles. Here we see the initial density profile $\rho_{0}(r)$ and current waveform $I(t)$ for $B_{z0} = 0$ shots 6457, 6462, 6467, 6474, and 6476, as well as $B_{z0} = 0.2\text{T}$ shots 6459, 6465, 6471, 6472, 6475, and 6477, as measured by planar laser-induced fluorescence (PLIF)~\cite{PLIF} and a Rogowski coil around the load, respectively.\label{Fig:Bz0-input}}
\end{figure}

\vspace{-5pt}
We also tested the model on two ensembles of highly repeatable shots previously analyzed in Lavine et al.~\cite{Lavine2024}. The first ensemble, consisting of shots 6457, 6462, 6467, 6474, and 6476, had no applied axial field. Its triple-nozzle initial density profile and current waveform are displayed in Fig.~\ref{Fig:Bz0-input}. The second ensemble, consisting of shots 6459, 6465, 6471, 6472, 6475, and 6477, was initially magnetized with uniform axial field $B_{z0} = 0.2\text{T}$. Its initial density profile and current waveform were approximately identical. The results for these ensembles are shown in Figs.~\ref{Fig:Bz0-0-r} and \ref{Fig:Bz0-0.2-r}, respectively. The model predictions once again agree closely with the data. However, further experimental comparisons are necessary to assess model fidelity for stronger axial fields.
\vspace{-5pt}
\subsection{iv. Numerical Implementation}
\vspace{-5pt}
We employed the forward Euler finite-difference method for all differentials. The structure of our code is modular with respect to each stage of the implosion model, with transitions occurring when appropriate conditions are met. Only the separation time $t_{s}$ is ambiguously defined, its minimum necessary value likely depending on the resolution and integration scheme of a given implementation. Fortunately, the results are relatively insensitive to its value, and we arbitrarily used $t_{s} = 1\text{ns}$ for all comparisons, with the inward stage commencing after this fixed time elapses.
\begin{figure}[h!tbp]
\centering
\noindent\includegraphics[width=0.965\columnwidth]{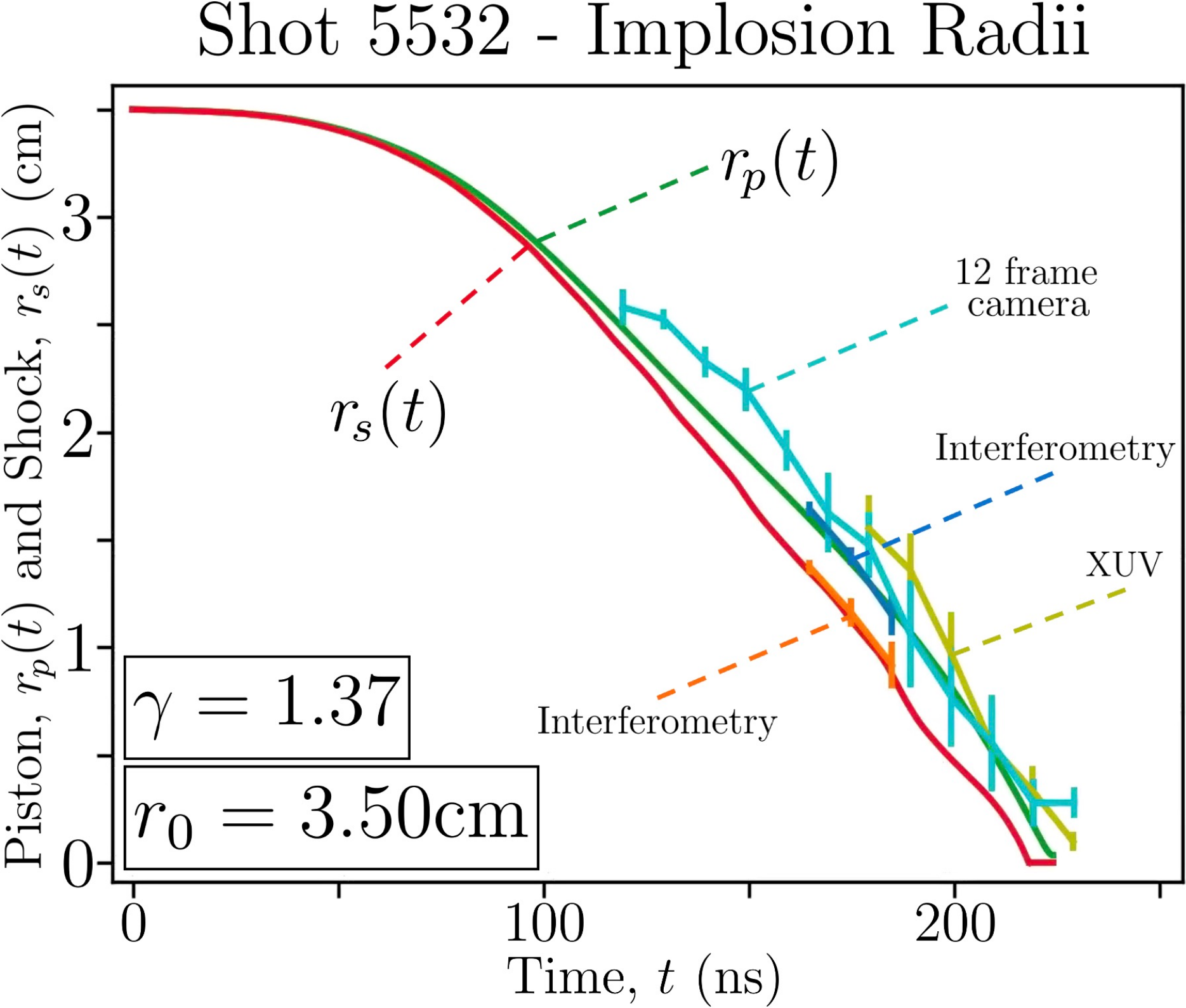}
\caption{Radial Trajectories Comparison for Shot 5532. Here we see a comparison of the calibrated trajectories of the piston and shock radii for shot 5532, together with various diagnostic measurements of their true values, after an MSE fit yielded optimal parameters $(r_{0},\gamma) = (3.50\text{cm},1.37)$.\label{Fig:5532-r}}
\end{figure}
\begin{figure}[h!tbp]
\centering
\noindent\includegraphics[width=\columnwidth]{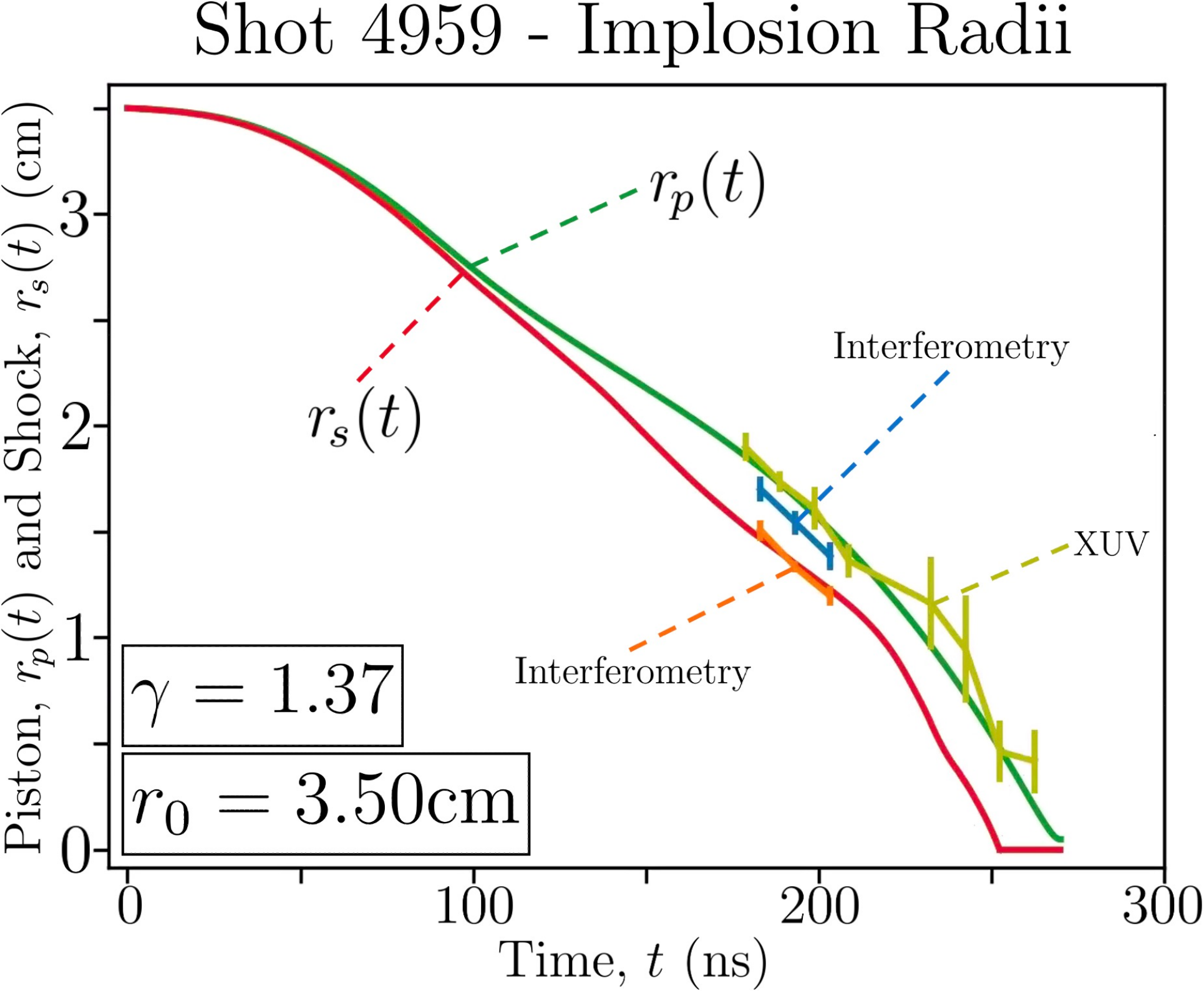}
\caption{Radial Trajectories Comparison for Shot 4959. Here we see a comparison of the predicted trajectories of the piston and shock radii for shot 4959, together with various diagnostic measurements, using calibrated parameters $(r_{0},\!\gamma) = (3.50\text{cm},1.37)$ based on MSE fitting to shot 5532.\label{Fig:4959-r}}
\end{figure}
\clearpage
\vspace{-25pt}
\subsection{v. Model Limitations}
\begin{figure}[h!tbp]
\centering
\noindent\includegraphics[width=\columnwidth]{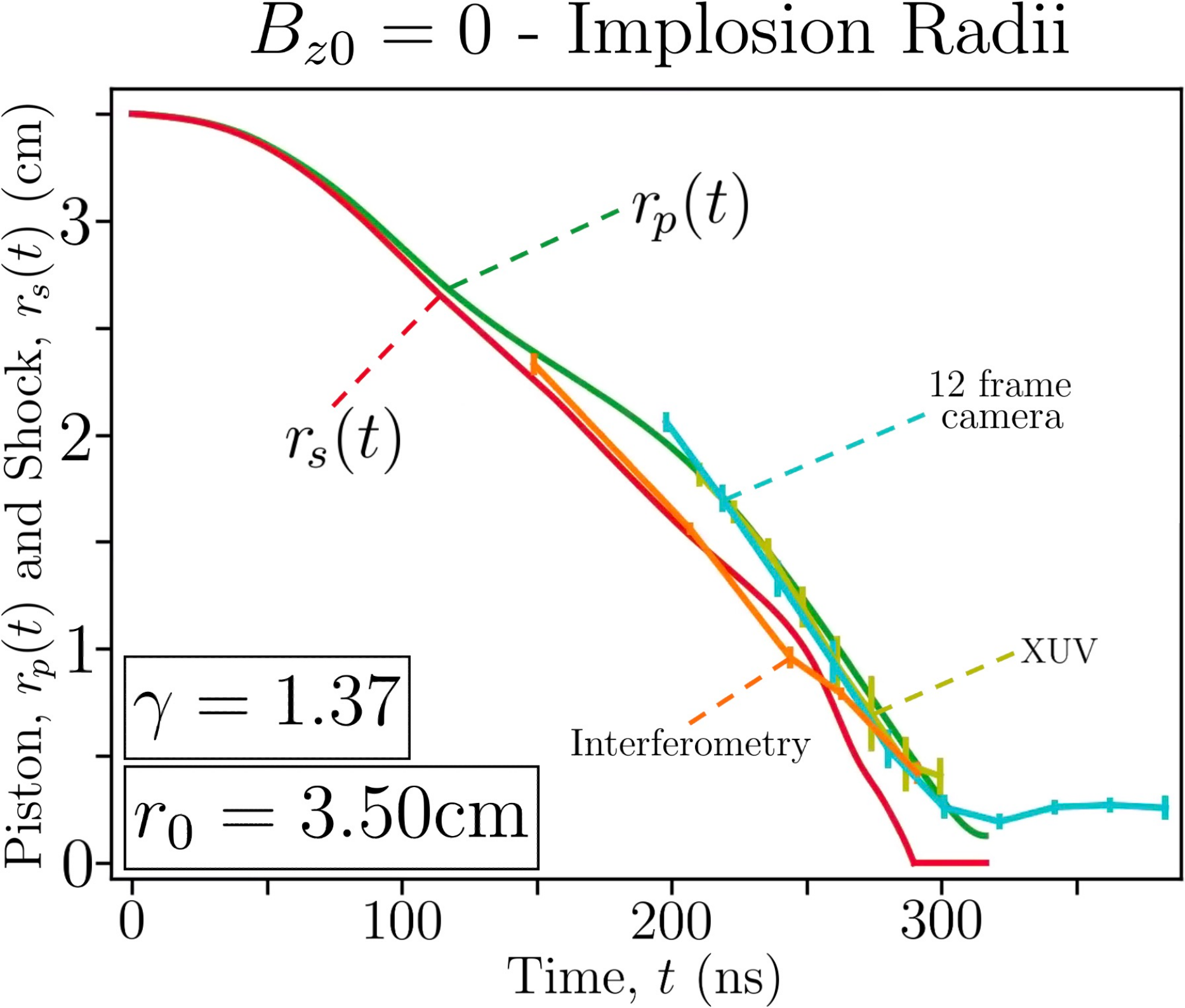}
\caption{Radial Trajectories Comparison for the $B_{z0} = 0$ Ensemble. Here we see a comparison of the predicted trajectories of the piston and shock radii for $B_{z0} = 0$ shots 6457, 6462, 6467, 6474, and 6476, along with various diagnostic measurements, using calibrated parameters $(r_{0},\!\gamma) = (3.50\text{cm},1.37)$ based on MSE fitting to shot 5532.\label{Fig:Bz0-0-r}}
\end{figure}

\begin{figure}[h!tbp]
\centering
\noindent\includegraphics[width=\columnwidth]{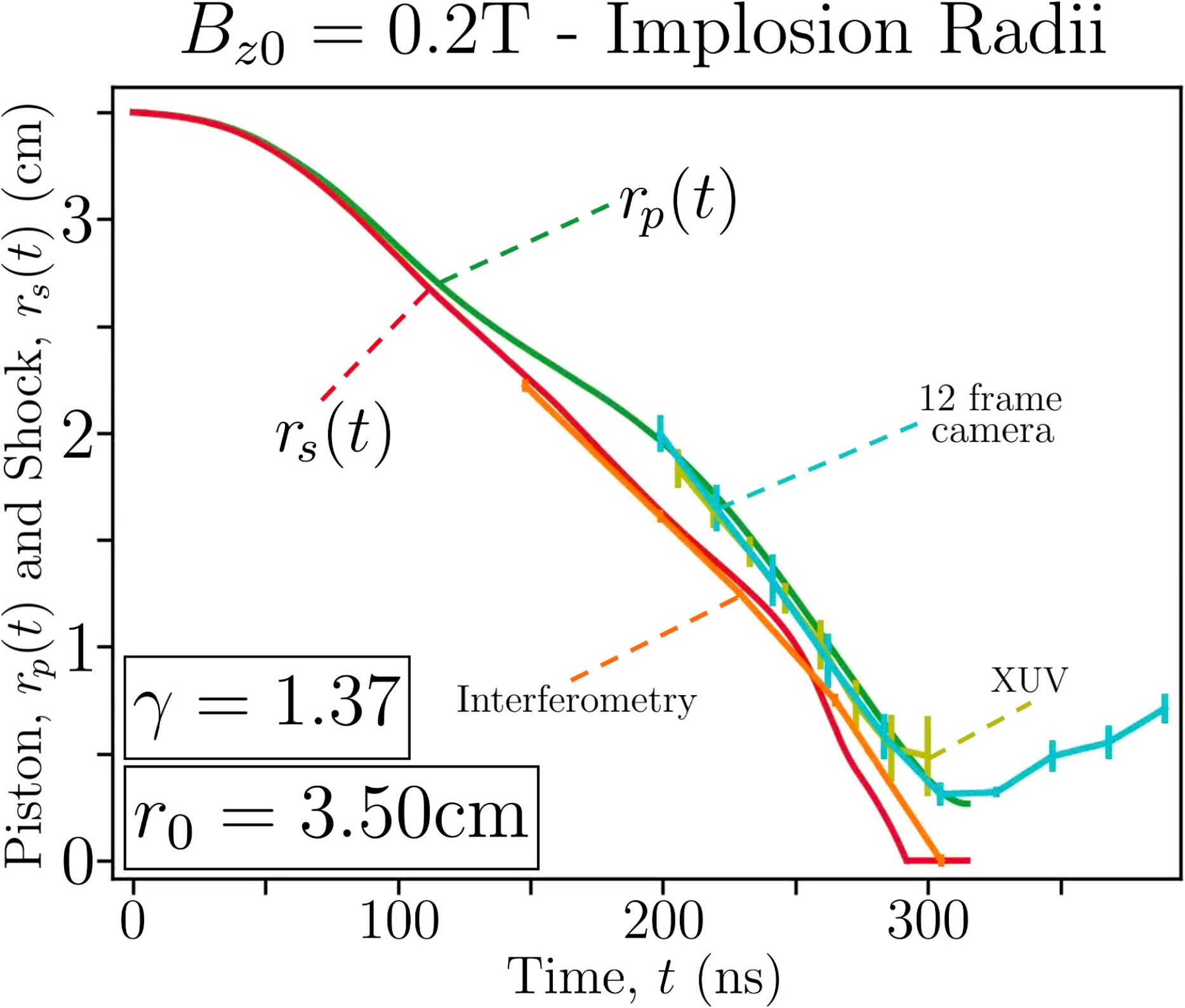}
\caption{Radial Trajectories Comparison for the $B_{z0} = 0.2\text{T}$ Ensemble. Here we see a comparison of the predicted trajectories of the piston and shock radii for $B_{z0} = 0.2\text{T}$ shots 6459, 6465, 6471, 6472, 6475, and 6477, along with various diagnostic measurements, using calibrated parameters $(r_{0},\!\gamma) = (3.50\text{cm},1.37)$ based on MSE fitting to shot 5532.\label{Fig:Bz0-0.2-r}}
\end{figure}

\vspace{-9pt}
While the results in Figs.~\ref{Fig:5532-r},\ref{Fig:4959-r},\ref{Fig:Bz0-0-r},\ref{Fig:Bz0-0.2-r} are encouraging, small discrepancies between model predictions and experimental observations remain. One source of error is violation of the uniform pressure profile assumption discussed in Appendix~\hyperref[C:I]{C.I}. Radiative losses throughout the implosion are also neglected, which, even prior to stagnation, might significantly affect the dynamics.\\
\indent Several physical effects may also contribute. For example, the magneto-Rayleigh-Taylor (MRT) instability~\cite{Qi2014,deGrouchy2018,Lavine2021} can drive axial transport, violating the model's assumption of translational symmetry. One consequence is zippering~\cite{Lavine2021}, where different heights implode at different rates, whereas the measurements reported here represent an axial average. Sausage and kink instabilities, turbulence, and departures from ideal MHD may further contribute to uncertainty.\\
\indent Additionally, emissive diagnostics (12-frame camera and XUV) may infer a larger apparent piston radius due to radiation from a surrounding low-density corona~\cite{Haines2011,Woolstrum2022}, while laser interferometry identifies a sharper density cutoff. Perfect agreement with experiments cannot reasonably be expected, making the overall correspondence all the more striking.
\section{VI. Sheath Profiles}
\label{sec:profiles}
\vspace{-5pt}
As described in Sec.\,\hyperref[sec:model]{IV}, we take the sound speed in the postshock region to be sufficiently large that its pressure profile is uniform. This should be approximately true for a strong inward shock, and scaling arguments are presented in Appendix~\hyperref[C:I]{C.I}. The velocity and pressure profiles can then be derived from the RH jump conditions in Appendix~\hyperref[App:A]{A} and the adiabatic law in Eq.~(\ref{Eq:MHD-3}) as shown in Appendices~\hyperref[App:B]{B} and~\hyperref[App:C]{C}, respectively.\\
\indent Upon recognizing that the shock is the only entropic agent, we can further assert that individual fluid elements should evolve adiabatically on either side of the shock front. That is, the $i$th fluid element should expand or compress with $P_{i}(t)/\rho_{i}(t)^{\gamma}$ held constant. 
Together with the pressure profiles and conservation of mass, this assumption is sufficient to determine all of their densities, radial widths, and motions. Taking the initial axial field to be frozen into the fluid elements, as dictated by Alfv\'en's theorem, is then sufficient to determine the axial field profile as well. These calculations require tracking the motions of all the individual fluid elements with a Lagrangian coordinate system and are performed in detail in Appendix~\hyperref[App:D]{D}.
\vspace{-10pt}
\subsection{i. Separation \& Inward Stages}
\vspace{-5pt}
One can derive the Eulerian velocity and pressure profiles in the postshock ($r_{s} < r < r_{p}$) region directly from the adiabatic law in Eq.~(\ref{Eq:MHD-3}) and the RH jump conditions as shown in Appendices~\hyperref[B:I]{B.I} and \hyperref[C:II]{C.II}, respectively.
\begin{align}\label{Eq:velocity}&v_{l}(r,t)=\!\frac{(r^{2}\!-\!r_{s}^{2})r_{p}\dot{r}_{p}\!+\!(r_{p}^{2}\!-\!r^{2})\frac{2}{\gamma+1}r_{s}\dot{r}_{s}}{(r_{p}^{2}\!-\!r_{s}^{2})r},\\
\label{Eq:pressure}
&P_{l}(t) = \frac{2}{\gamma+1}\rho_{0}(r_{s})\dot{r}_{s}^{2}.\end{align}
Determining the density and axial field profiles requires tracking the individual fluid elements in a Lagrangian frame as shown in Appendix~\hyperref[D:II]{D.II}.\\
\indent Let us define $t_{si}$ to be the time when the shock front first encounters the $i$th fluid element. In the untouched ($0\leq r < r_{s}$) region, the initial pressure, density, axial field, position, radial width, and velocity of the $i$th fluid element, that is, for $0\leq t < t_{si}$, are simply,
\begin{align}
&P_{i}(t)\!=\! P_{0}(r_{i}(0)),\\
&\rho_{i}(t) \!=\! \rho_{0}(r_{i}(0)\!),\\
&B_{zi}(t)\!=\! B_{z0},\\
&r_{i}(t) \!=\! r_{i}(0),\\
&dr_{i}(t) \!=\! dr_{i}(0),\\
&v_{i}(t) = 0.\end{align}
The pressure, density, axial field, position, radial width, and velocity of the $i$th fluid element in the postshock ($r_{s} < r < r_{p}$) region, that is, for $t_{si} < t < t_{c}$, are then,
\begin{align}
\label{Eq:Pi}
&P_{i}(t) = P_{l}(t),\\
\label{Eq:density}
&\rho_{i}(t) = \frac{\gamma+1}{\gamma-1}\!\left(\!\frac{P_{l}(t)}{P_{l}(t_{si})}\!\right)^{\!\!1/\gamma}\!\!\!\!\!\!\!\rho_{0}(r_{i}(0)\!),\\
\label{Eq:axial}
&B_{zi}(t) = B_{z0}\frac{\rho_{i}(t)}{\rho_{i}(0)},\end{align}
\begin{align}
\label{Eq:position}
&r_{i}(t) = \sqrt{r_{p}(t)^{2}\!-\!(r_{p}(t_{si})^{2}\!-\!r_{i}(0)^{2})\!\!\left(\!\frac{P_{l}(t_{si})}{P_{l}(t)}\!\right)^{\!\!1/\gamma}\!},\\
\label{Eq:width}
&dr_{i}(t) = \frac{r_{i}(0)\rho_{i}(0)}{r_{i}(t)\rho_{i}(t)}dr_{i}(0),\\
\label{Eq:vi}
&v_{i}(t) = \dot{r}_{i}(t) = v_{l}(r_{i}(t),t).
\end{align}
Using the positions in Eq.~(\ref{Eq:position}), the Lagrangian profiles in Eqs.~(\ref{Eq:Pi}-\ref{Eq:vi}) can be used to reconstruct the Eulerian profiles through numerical interpolation, namely $\rho_{l}(r_{i}(t),t) = \rho_{i}(t)$ and $B_{zl}(r_{i}(t),t) = B_{zi}(t)$.
\vspace{-10pt}
\subsection{ii. Compression Stage}
\vspace{-5pt}
We can once again derive the Eulerian velocity and pressure profiles in the postshock ($0 \leq r < r_{p}$) region directly from the adiabatic law in Eq.~(\ref{Eq:MHD-3}) and the RH jump conditions as shown in Appendices~\hyperref[B:II]{B.II} and \hyperref[C:III]{C.III}, respectively.
\begin{align}\label{Eq:velocity-compression}&v_{l}(r,t) = \frac{r\dot{r}_{p}}{r_{p}},\\
\label{Eq:pressure-compression}&P_{l}(t) = \frac{2}{\gamma\!+\!1}\rho_{0}(0)\dot{r}_{s}(t_{c}^{-})^{2}\!\left(\!\frac{r_{p}(t_{c})}{r_{p}(t)}\!\right)^{\!\!2\gamma}\!\!\!\!\!.\end{align}
The on-axis boundary condition of the velocity profile in the postshock region in Eq.~(\ref{Eq:velocity-compression}) appears to disagree with that of Eq.~(\ref{Eq:velocity}) at the moment of reflection. This abrupt transition can be understood via the formation of a boundary layer that resolves at the moment of incidence, as illustrated in Appendix~\hyperref[B:III]{B.III}, and energy is indeed conserved, as shown in Appendix~\hyperref[B:IV]{B.IV}. The pressure profile in Eq.~(\ref{Eq:pressure-compression}) merely states that the entire column enclosed by the piston is being adiabatically compressed. Determining the density and axial field profiles requires tracking the individual fluid elements in a Lagrangian frame as shown in Appendix~\hyperref[D:III]{D.III}.\\
\indent The pressure, density, axial field, position, radial width, and velocity of the $i$th fluid element in the postshock $(0 \leq r < r_{p})$ region, that is, for $t_{c} \leq t \leq t_{p}$, are thus,
\begin{align}
\label{Eq:Pi-compression}
&P_{i}(t) = P_{l}(t),\\
\label{Eq:density-compression}
&\rho_{i}(t) = \rho_{i}(t_{c})\frac{r_{p}(t_{c})^{2}}{r_{p}(t)^{2}},\\
\label{Eq:axial-compression}&B_{zi}(t) = B_{z0}\frac{\rho_{i}(t)}{\rho_{i}(0)},\\
\label{Eq:position-compression}
&r_{i}(t) = \frac{r_{p}(t)}{r_{p}(t_{c})}r_{i}(t_{c}),\\
\label{Eq:width-compression}
&dr_{i}(t) = \frac{r_{p}(t)}{r_{p}(t_{c})}dr_{i}(t_{c}),\\
\label{Eq:vi-compression}
&v_{i}(t) = \dot{r}_{i}(t) = \frac{r_{i}(t_{c})}{r_{p}(t)}\dot{r}_{p}(t).\end{align}
\begin{figure}
\centering
\noindent\includegraphics[width=\columnwidth]{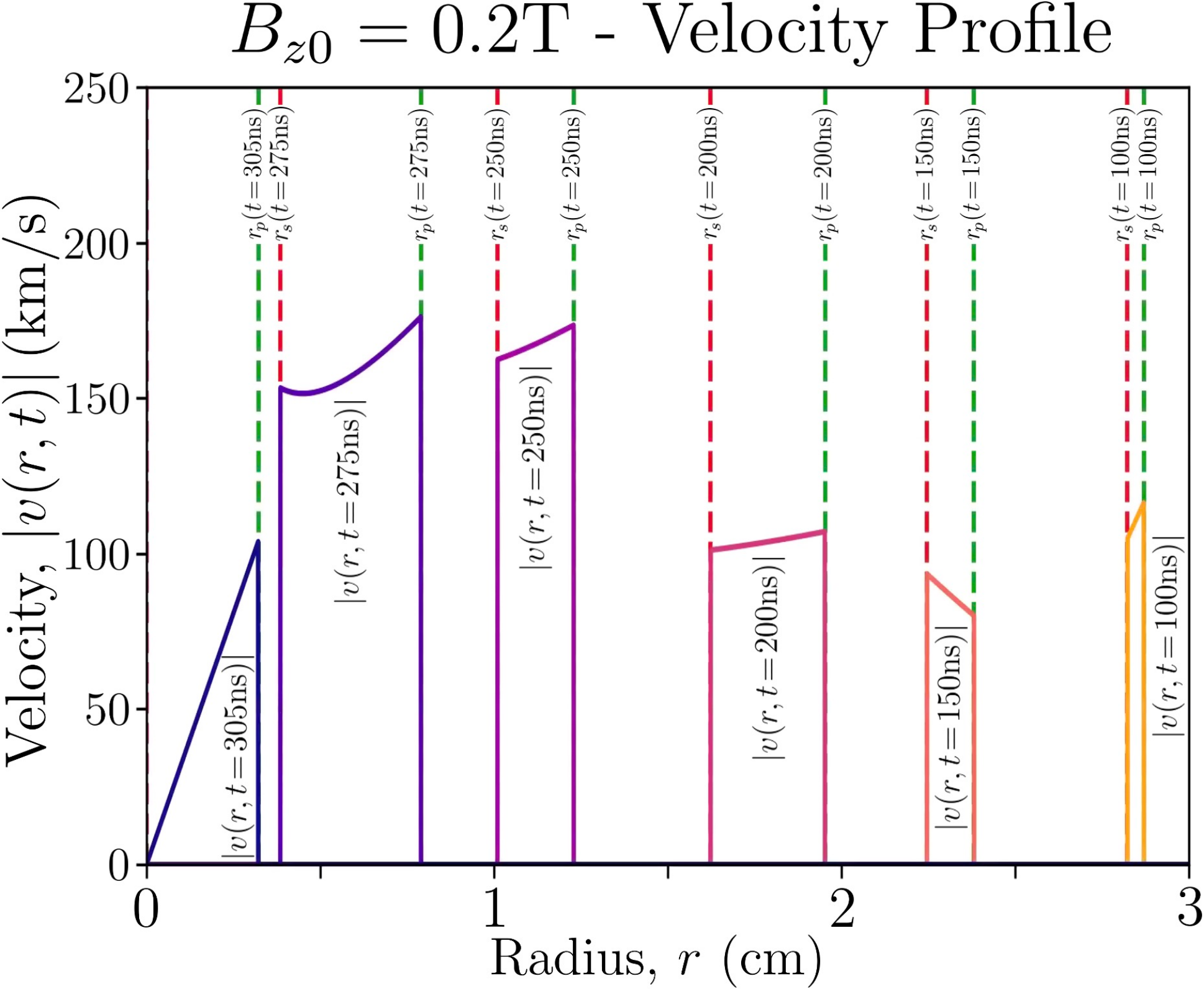}
\caption{Velocity Profiles $v(r,t)$ for the $B_{z0} = 0.2$T Ensemble. Here we see the predicted radial flow velocity profiles for $B_{z0} = 0.2$T shots 6459, 6465, 6471, 6472, 6475, and 6477 for times $t = \{100\text{ns},150\text{ns},200\text{ns},250\text{ns},275\text{ns},305\text{ns}\}$. The boundary layer behavior described in Appendix~\hyperref[B:III]{B.III} occurs between $t = 275\text{ns}$ and $t = 305\text{ns}$.\label{Fig:4959-v}}
\end{figure}
\begin{figure}
\centering
\noindent\includegraphics[width=\columnwidth]{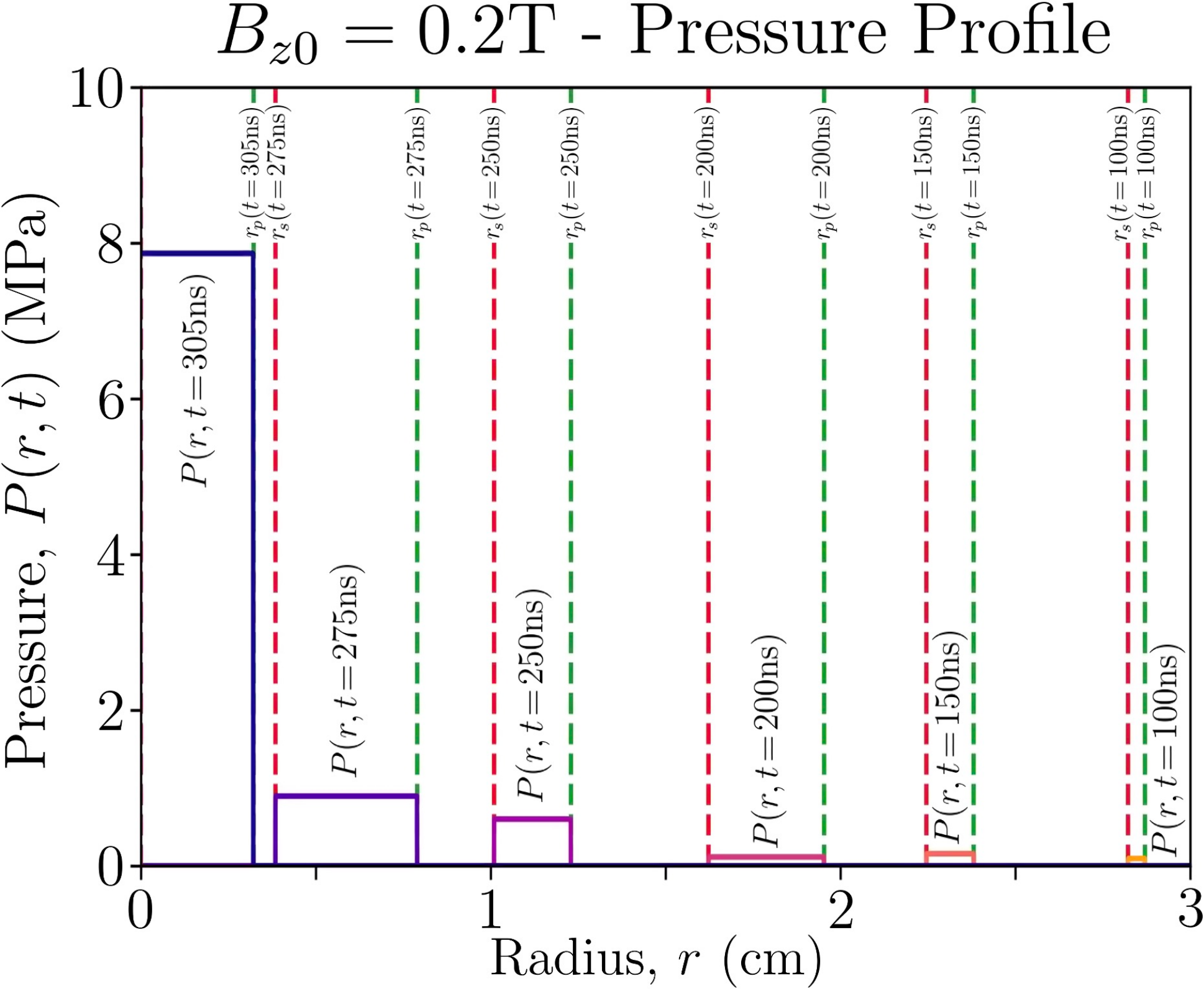}
\caption{Pressure Profiles $P(r,t)$ for the $B_{z0} = 0.2$T Ensemble. Here we see the predicted pressure profiles for $B_{z0} = 0.2$T shots 6459, 6465, 6471, 6472, 6475, and 6477 for times $t = \{100\text{ns},150\text{ns},200\text{ns},250\text{ns},275\text{ns},305\text{ns}\}$. Evidently, these uniform profiles are an oversimplification, but the steep evolution in magnitude over time is worth noting.\label{Fig:4959-p}}
\end{figure}
\begin{figure}
\centering
\noindent\includegraphics[width=\columnwidth]{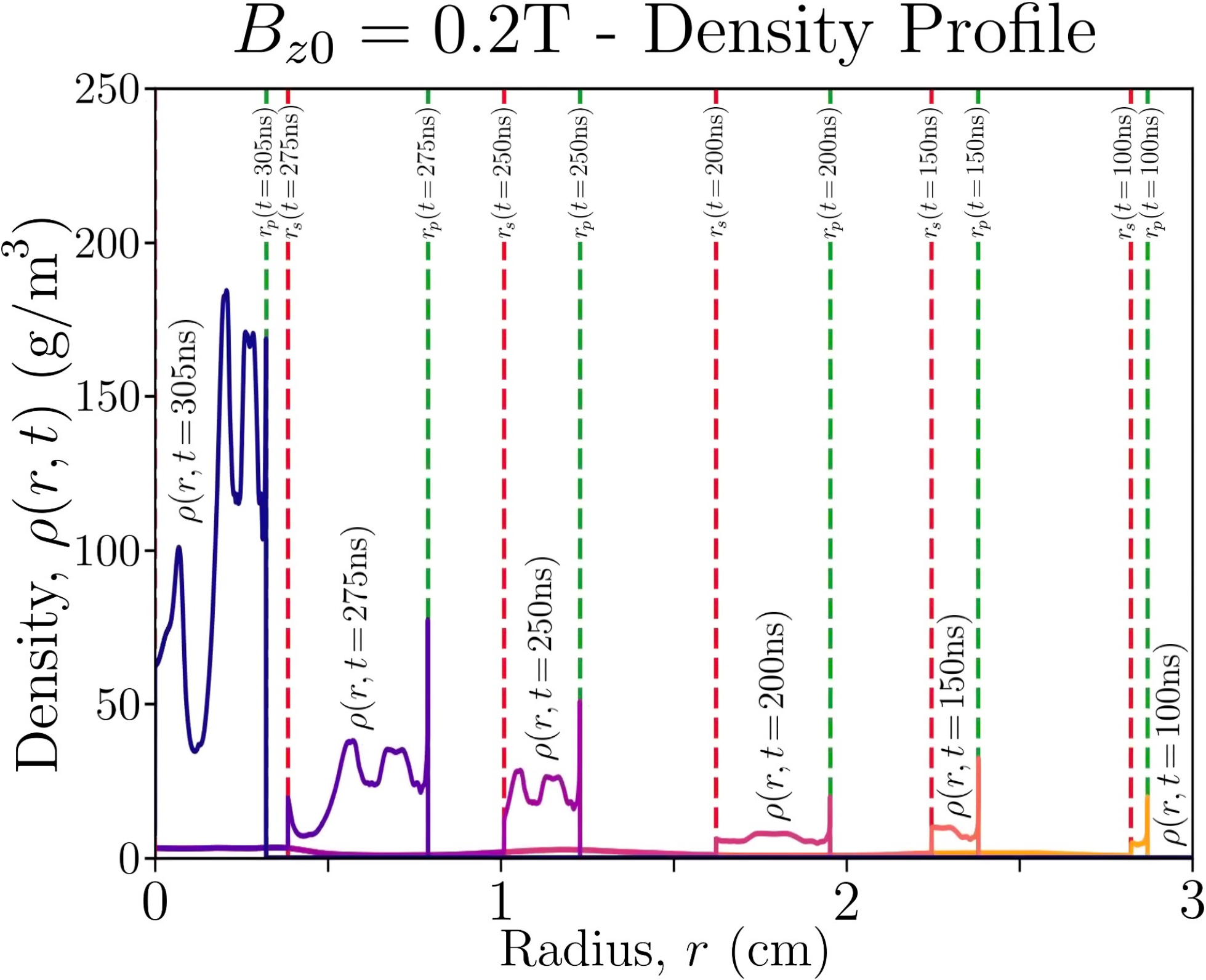}
\caption{Density Profiles $\rho(r,t)$ for the $B_{z0} = 0.2$T Ensemble. Here we see the predicted density profiles for $B_{z0} = 0.2$T shots 6459, 6465, 6471, 6472, 6475, and 6477 for times $t = \{100\text{ns},150\text{ns},200\text{ns},250\text{ns},275\text{ns},305\text{ns}\}$. We observe a clear compressed reproduction of the initial density profile within the sheath.\label{Fig:4959-rho}}
\end{figure}
\begin{figure}
\centering
\noindent\includegraphics[width=\columnwidth]{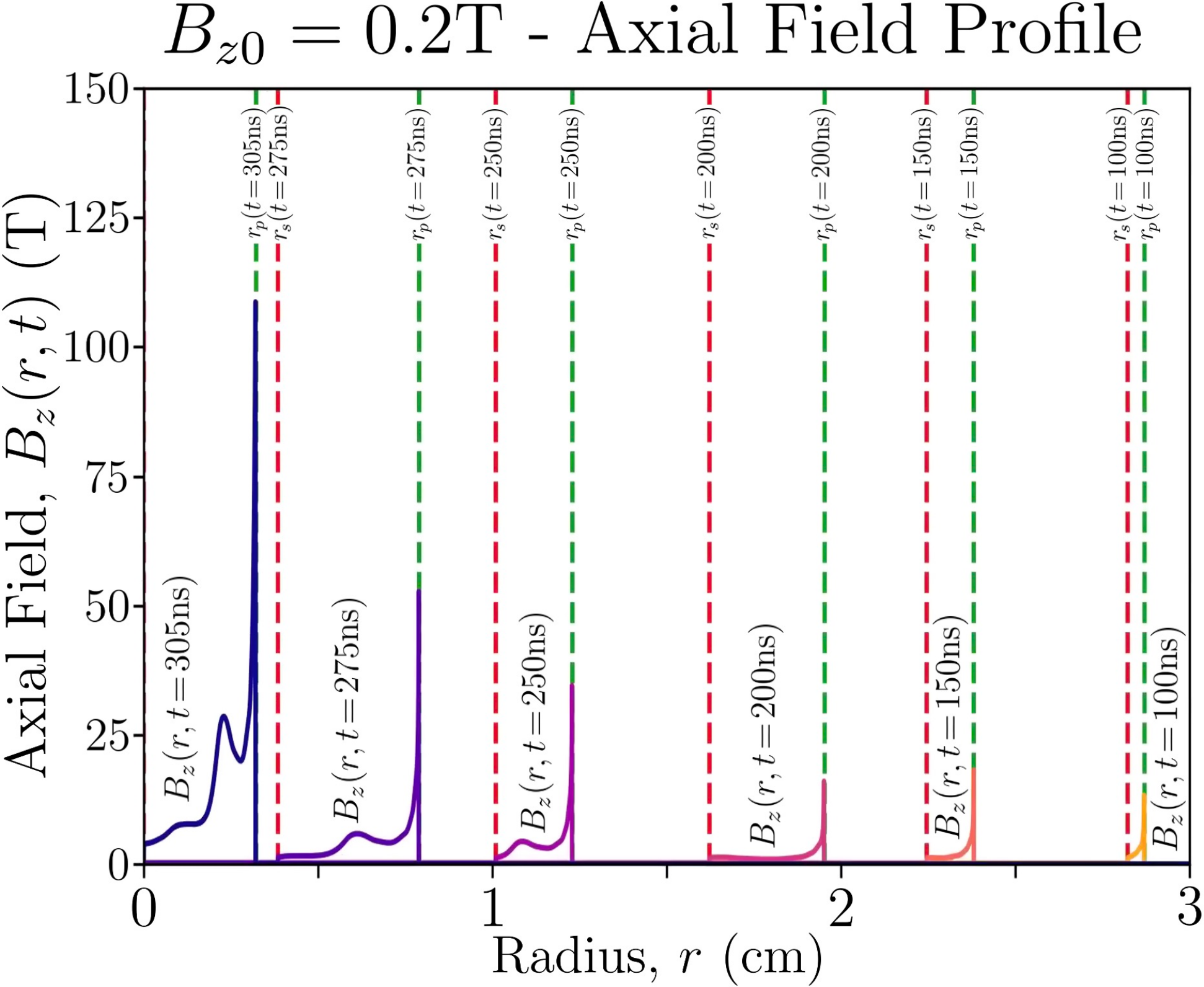}
\caption{Axial Field Profiles $B_{z}(r,t)$ for the $B_{z0} = 0.2$T Ensemble. Here we see the predicted axial field profiles for $B_{z0} = 0.2$T shots 6459, 6465, 6471, 6472, 6475, and 6477 for times $t = \{100\text{ns},150\text{ns},200\text{ns},250\text{ns},275\text{ns},305\text{ns}\}$. We observe a sharply peaked axial field near the piston.\label{Fig:4959-Bz}}
\end{figure}\newpage
\vspace{-10pt}
\indent The Eulerian velocity and pressure profiles from Eqs.~(\ref{Eq:velocity}-\ref{Eq:pressure}) and Eqs.~(\ref{Eq:velocity-compression}-\ref{Eq:pressure-compression}), as well as the interpolated Lagrangian profiles from Eqs.~(\ref{Eq:Pi}-\ref{Eq:vi}) and Eqs.~(\ref{Eq:Pi-compression}-\ref{Eq:vi-compression}), are plotted at various moments throughout the implosion in Figs.~\ref{Fig:4959-v},\ref{Fig:4959-p},\ref{Fig:4959-rho},\ref{Fig:4959-Bz} for the $B_{z0}=0.2$T ensemble.\\
\indent Although no experimental data are available for comparison, we observe a clear compressed reproduction of the initial density profile within the sheath, an axial field profile concentrated near the piston, and a rapidly increasing pressure profile as the shock reaches the axis. Apart from the localized divergence of the density and axial field profiles at the piston due to the unphysical singular point of the adiabatic compression model in Eq.~(\ref{Eq:density}) at the initial conditions in Eq.~(\ref{Eq:IC}), these profiles appear physically reasonable. This divergence essentially stems from our simplifying assumption of uniform pressure profile within the sheath, and the resulting discontinuity in the net material and magnetic pressure across the infinitely dense piston is absolutely necessary to capture its acceleration under this simplifying assumption.\\
\indent We can also calculate the maximum on-axis values of the density and pressure at the moment of stagnation from Eqs.~(\ref{Eq:pressure-compression},\ref{Eq:density-compression}),
\begin{align}&\tilde{\rho} \!=\!\frac{\gamma\!+\!1}{\gamma\!-\!1}\rho_{0}(0)\frac{r_{p}(t_{c})^{2}}{r_{p}(t_{p})^{2}},\\
&\tilde{P}\!=\! \frac{2}{\gamma\!+\!1}\rho_{0}(0)\dot{r}_{s}(t_{c}^{-})^{2}\!\!\left(\!\frac{r_{p}(t_{c})}{r_{p}(t_{p})}\!\right)^{\!\!2\gamma}.
\end{align}
\vspace{-15pt}
\section{VII. Fundamental Relation}
\begin{figure}
\centering
\noindent\includegraphics[width=0.95\columnwidth,height=0.8\columnwidth]{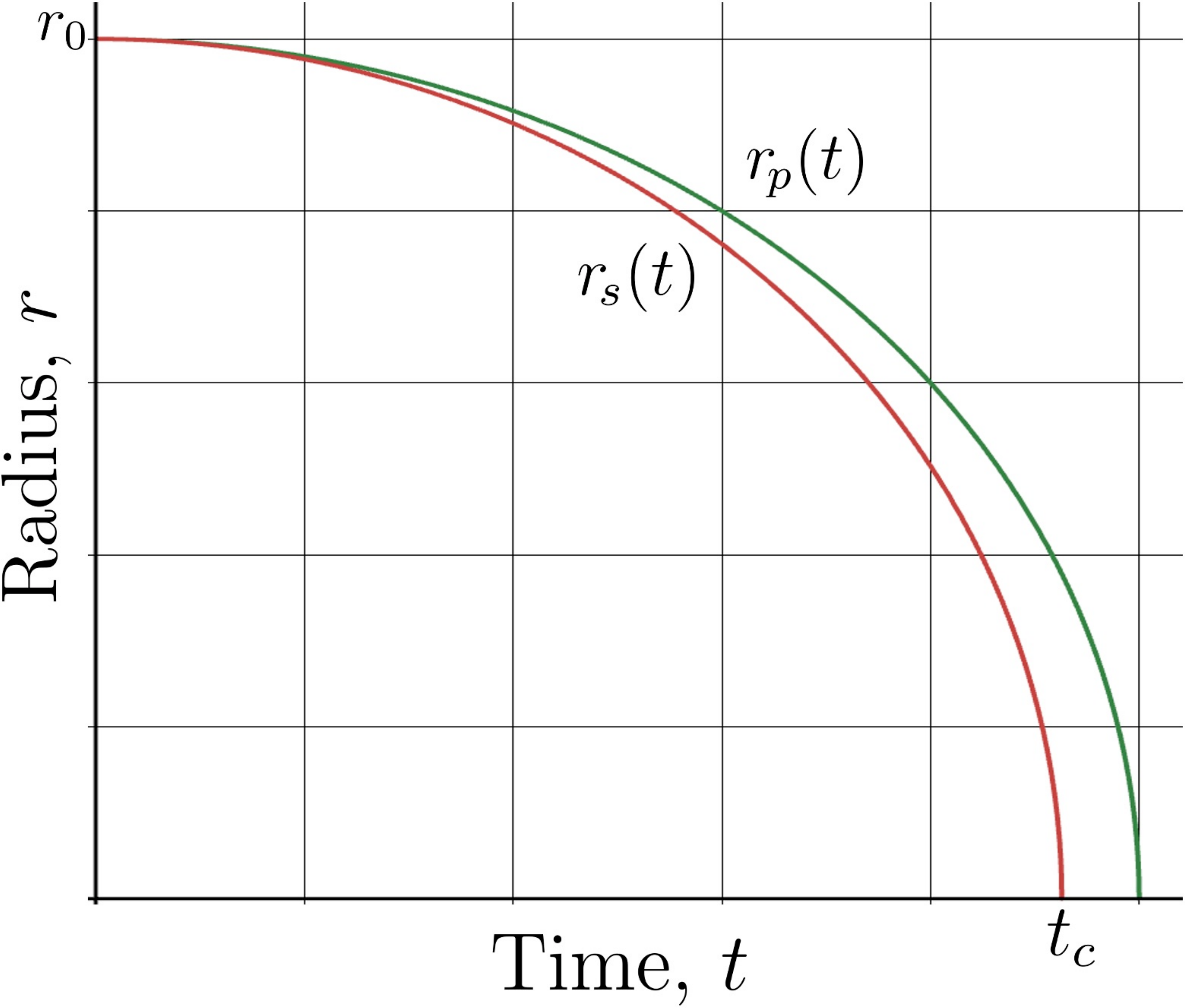}
\caption{Geometric Relation. Shown here is an exemplary solution of the simple fundamental relation between the piston and shock (Eq.~(\ref{Eq:geometry})\!) that lays bare the essence of our model. When the additional structure and degrees of freedom resulting from finite sheath thickness and spatiotemporal dependence of our variables fall away, this is the basic coupling that remains. Here $t_{c}$ is the incidence time of the shock on axis.\label{Fig:geometry}}
\end{figure}

\vspace{-5pt}
\label{sec:fundamental}
\indent Finally, we can extract an interesting approximate geometric relationship between the shock and piston radii from the dominant balance of Potter's model in Eq.~(\ref{Eq:Potter}),
\begin{align}\label{Eq:geometry}
r_{p}\dot{r}_{p} = \frac{2}{\gamma+1}r_{s}\dot{r}_{s}\,\longleftrightarrow\, r_{p}^{2}\!-\!r_{s}^{2} = \!\frac{\gamma\!-\!1}{\gamma\!+\!1}(r_{0}^{2}\!-\!r_{s}^{2}).\end{align}
For example, if $r_{s}(t) = r_{0}\sqrt{1-t^{2}/t_{c}^{2}}$ where $t_{c}$ is the incidence time, then $r_{p}(t) = r_{0}\sqrt{1-\frac{2}{\gamma+1}\frac{t^{2}}{t_{c}^{2}}}$. These represent reasonable, compatible physical trajectories as shown in Fig.~\ref{Fig:geometry}. Eq.~(\ref{Eq:geometry}) is quite useful for making quick estimates of the relative positions of the piston and shock, and can be viewed 
as the fundamental relationship between them. This idealization stands until modified by the changing current, nonconstant density, axial field pressure, and finite thickness of the sheath, in the ways we have presented here.\\
\indent Further corrections beyond those we have included, such as spatially nonuniform pressure profiles, a nonuniform initial axial field, or relaxation of the strong inward shock assumption together with consideration of an initial pressure profile, can be easily amended as necessary without altering our general framework. No matter the assumptions or regime, the trajectories of the piston and shock can still be described via a set of coupled ODEs obtained directly from the ideal MHD momentum and adiabatic equations integrated over some limited domain. The velocity and pressure profiles in various regions can be subsequently derived from the RH jump conditions and the adiabatic law. Everything else, including the motions of the fluid elements, and the density and axial field profiles, as well as the implosion time and stagnation radius, are then fully determined by conservation of mass, momentum, energy, and magnetic flux.
\vspace{-10pt}
\section{VIII. Conclusions}
\label{sec:conclusions}
\vspace{-5pt}
This multi-stage analytical approach to z-pinch modeling is insightful, simple, and agrees with observations. Our model predicts the radial implosion trajectories of the piston and shock with remarkable accuracy and additionally describes the fluid element trajectories as well as the velocity, pressure, density, and axial field profiles in all regions throughout the implosion. This is the first analytical model of the z-pinch to incorporate an axial magnetic field in addition to spatially varying initial density and time-dependent axial current for a sheath of finite thickness. This is also the first detailed comparison of an analytical model with experimental data from COBRA and represents a significant step forward in our physical understanding of the dynamic z-pinch.
Our model should provide a strong basis for quickly analyzing and designing z-pinch experiments at pulsed-power facilities worldwide.
\vspace{-15pt}
\section{Acknowledgments}
\vspace{-5pt}
We would like to thank the entire Laboratory of Plasma Studies staff for recording and providing the experimental data used for comparison with our model as well as H. Fetsch, N. S. Chopra, T. Foster, P. Karypis, W. Chu, and the reviewer for their absolutely invaluable insight throughout the development of this paper.\\
\indent This material is based upon work supported by the U.S. Department of Energy under Grant No. DE-AC02-09CH11466, and the National Nuclear Security Administration's Stewardship Sciences Academic Programs under Grants No. DE-NA0003746 and No. DE-NA0004148.
\bibliography{zpinch}
\bibliographystyle{abbrv}
\bibliographystyle{unsrt}
\newpage
\section{Appendix A: Rankine-Hugoniot Jump Conditions}
\label{App:A}
\renewcommand{\theequation}{A\arabic{equation}}
\renewcommand{\theHequation}{A\arabic{equation}}
\setcounter{equation}{0}
The Rankine-Hugoniot (RH) jump conditions~\cite{Classic,Zeldovich} are relations describing changes in the density $\rho$, flow velocity (in the shock frame) $\Vec{u}$, pressure $P$, and magnetic field $\Vec{B}$ across a shock front. A shock front is a propagating wave that moves faster than the local speed of sound of the unperturbed fluid and is characterized by abrupt, essentially discontinuous changes in behavior across it.\\
\indent In ideal MHD, there are five basic RH jump conditions, three from conservation of mass, momentum, and energy across the front, and two from the fundamental boundary conditions imposed by Maxwell's equations. Using subscripts $1$ and $2$ to denote the upstream (initial) and downstream (postshock) quantities, respectively, let us define the square bracket $[a] \equiv a_{2} - a_{1}$. Then across a shock front characterized by normal vector $\mathbf{\hat{n}}$, the jump conditions in the shock frame are given as follows,
\begin{align}
\label{RH:mass}
&\left[\rho\vec{u}\cdot\mathbf{\hat{n}}\right] = 0,\\
\label{RH:momentum}
&\left[\!\left(\!\rho\vec{u}\vec{u}+\!\left(\!P+\frac{B^{2}}{2\mu_{0}}\right)\!\overline{\overline{{1}}}-\frac{\vec{B}\vec{B}}
{\mu_{0}}\right)\!\cdot\mathbf{\hat{n}}\right] = 0,\\
\label{RH:energy}
&\left[\!\left(\!\!\left(\frac{1}{2}\rho u^{2}+\frac{\gamma}{\gamma-1}P\!\right)\!\vec{u}+\frac{\vec{E}\times\vec{B}}{\mu_{0}}\right)\!\cdot\mathbf{\hat{n}}\right] = 0,\\
\label{RH:e-field}
&\left[\vec{E}\times\mathbf{\hat{n}}\right] = \left[(\vec{B}\times\vec{u}\,)\times\mathbf{\hat{n}}\right] = 0,\\
\label{RH:b-field}
&\left[\vec{B}\cdot\mathbf{\hat{n}}\right] = 0.
\end{align}
We consider the case of a z-pinch with axial symmetry, and restrict the flow velocity and magnetic field on either side of the shock front to the following components in standard cylindrical coordinates.
\begin{align}\vec{u} = u_{r}\mathbf{\hat{r}},\quad\vec{B} = B_{z}\mathbf{\hat{z}}.\end{align}
Let us define the jump ratio,
\begin{align}\label{21:mass}\rho_{2}/\rho_{1}\equiv X.\end{align}
Then from Eq.~(\ref{RH:mass}),
\begin{align}\label{21:velocity}\left[\rho\vec{u}\cdot\mathbf{\hat{r}}\right]=\left[\rho u_{r}\right] = 0 \rightarrow u_{r2}/u_{r1}\equiv1/X.\end{align}
And from $-\mathbf{\hat{z}}\,\cdot\,$Eq.~(\ref{RH:e-field}),
\begin{align}\label{21:efield}\!\!-\mathbf{\hat{z}}\cdot\left[\left(B_{z}\mathbf{\hat{z}}\!\times\!u_{r}\mathbf{\hat{r}}\right)\!\times\!\mathbf{\hat{r}}\right]\!=\!\left[B_{z}u_{r}\right] \!=\! 0\rightarrow B_{z2}/B_{z1}\!=X.\!\!\end{align}
Then from $\mathbf{\hat{r}}\,\cdot\,$Eq.~(\ref{RH:momentum}),
\begin{align}\left[\rho u_{r}^{2}+P+\frac{B^{2}}{2\mu_{0}}\right] = 0.\end{align}
Using our definition of the bracket,
\begin{align}\label{21:momentum}\rho_{2}u_{r2}^{2}+P_{2}+\frac{B_{z2}^{2}}{2\mu_{0}} = \rho_{1}u_{r1}^{2}+P_{1}+\frac{B_{z1}^{2}}{2\mu_{0}}.\end{align}
Dividing through by $P_{1}$, we rewrite Eq.~(\ref{21:momentum}),
\begin{align}\label{21:momentum-2}\!\!\!\!\!\frac{\rho_{2}}{\rho_{1}}\frac{u_{r2}^{2}}{u_{r1}^{2}}\frac{\rho_{1}u_{r1}^{2}}{P_{1}}\!+\!\frac{P_{2}}{P_{1}}\!+\!\frac{B_{z1}^{2}}{2\mu_{0}P_{1}}\frac{B_{z2}^{2}}{B_{z1}^{2}} \!=\! \frac{\rho_{1}u_{r1}^{2}}{P_{1}}\!+\!1\!+\!\frac{B_{z1}^{2}}{2\mu_{0}P_{1}}.\!\!\!\end{align}
Let us define the upstream Mach number $M_{1}$ and plasma beta $\beta_{1}$ as follows,
\begin{align}\label{RH:mach}M_{1}^{2}\equiv\frac{\rho_{1}u_{r1}^{2}}{\gamma P_{1}},\,\,\beta_{1}\equiv\frac{P_{1}}{B_{1}^{2}/2\mu_{0}}.\end{align}
Then, using these definitions and rearranging Eq.~(\ref{21:momentum-2}),
\begin{align}\label{21:momentum-3}\frac{P_{2}}{P_{1}} \!=\! 1+\gamma M_{1}^{2}\!\left(\!1\!-\!\frac{1}{X}\!\right)\!+\!\frac{1\!-\!X^{2}}{\beta_{1}}.\end{align}
Then from Eq.~(\ref{RH:energy}),
\begin{align}\label{RH:energy-2}\left[\!\left(\frac{1}{2}\rho u_{r}^{2}\!+\!\frac{\gamma}{\gamma\!-\!1}P\!\right)\!u_{r}+\frac{B_{z}^{2}u_{r}}{\mu_{0}}\right] = 0.\end{align}
Using our definition of the bracket and dividing through by $\gamma u_{r2}P_{1}/(\gamma-1)$, we rewrite Eq.~(\ref{RH:energy-2}),
\begin{align}\label{21:energy}
&\frac{\gamma-1}{2}\frac{\rho_{2}u_{r2}^{2}}{\rho_{1}u_{r1}^{2}}\frac{\rho_{1}u_{r1}^{2}}{\gamma P_{1}}\!+\!\frac{P_{2}}{P_{1}}\!+\!2\frac{\gamma\!-\!1}{\gamma}\frac{B_{z2}^{2}}{B_{z1}^{2}}\frac{B_{z1}^{2}}{2\mu_{0}P_{1}}\nonumber\\
&=\!\left(\!\frac{\gamma-1}{2}\frac{\rho_{1}u_{r1}^{2}}{\gamma P_{1}}\!+\!1\!\right)\!\frac{u_{r1}}{u_{r2}}\!+\!2\frac{\gamma\!-\!1}{\gamma}\frac{B_{z1}^{2}}{2\mu_{0}P_{1}}\frac{u_{r1}}{u_{r2}}.
\end{align}
Using the definitions in Eq.~(\ref{RH:mach}) and rearranging terms,
\begin{align}\label{21:energy-2}
&\frac{P_{2}}{P_{1}}\!=\!X\!+\!\frac{\gamma-1}{2}M_{1}^{2}\!\left(\!X\!-\!\frac{1}{X}\!\right)\!+\!2\frac{\gamma-1}{\gamma}\frac{X(1\!-\!X)}{\beta_{1}}.\!\!
\end{align}
Now let us equate our two expressions for $P_{2}/P_{1}$ in Eqs.~(\ref{21:momentum-3},\ref{21:energy-2}) to obtain the following equation in $X$,
\begin{align}\label{RH:Quadratic}&(X\!-\!1)\!\bigl(AX^{2}\!+\!BX\!-C\bigr)\!\!=\!0,\\
&A\!\equiv\!2(2-\gamma),\\
&B\!\equiv\!\gamma(2(1\!+\!\beta_{1}\!)\!+\!(\gamma\!-\!1)\beta_{1}M_{1}^{2}),\\
&C\!\equiv\!\gamma(\gamma\!+\!1)\beta_{1}M_{1}^{2},\end{align}
where we can identify the trivial solution $X = 1$, which corresponds to no shock front at all, and the nontrivial solution,
\begin{align}\label{Eq:nontrivial}&X = \frac{1}{4(2-\gamma)}\biggl(\!\!-\gamma(2(1\!+\!\beta_{1}\!)\!+\!(\gamma\!-\!1)\beta_{1}M_{1}^{2})\\
&\pm\!\sqrt{\gamma^{2}(2(1\!+\!\beta_{1}\!)\!+\!(\gamma\!-\!1)\beta_{1}M_{1}^{2})^{2}\!+\!8\gamma(2\!-\!\gamma)(\gamma\!+\!1)\beta_{1}M_{1}^{2}}\biggr).\nonumber\end{align}
Only the `+' branch is physical. We now explore the strong shock limit of this solution, in accordance with standard asymptotic theory~\cite{Bender}.
\subsection{I. Separation \& Inward Stages}
\label{A:I}
\begin{figure}[h]
\centering
\noindent\includegraphics[width=\columnwidth,height=0.5\columnwidth]{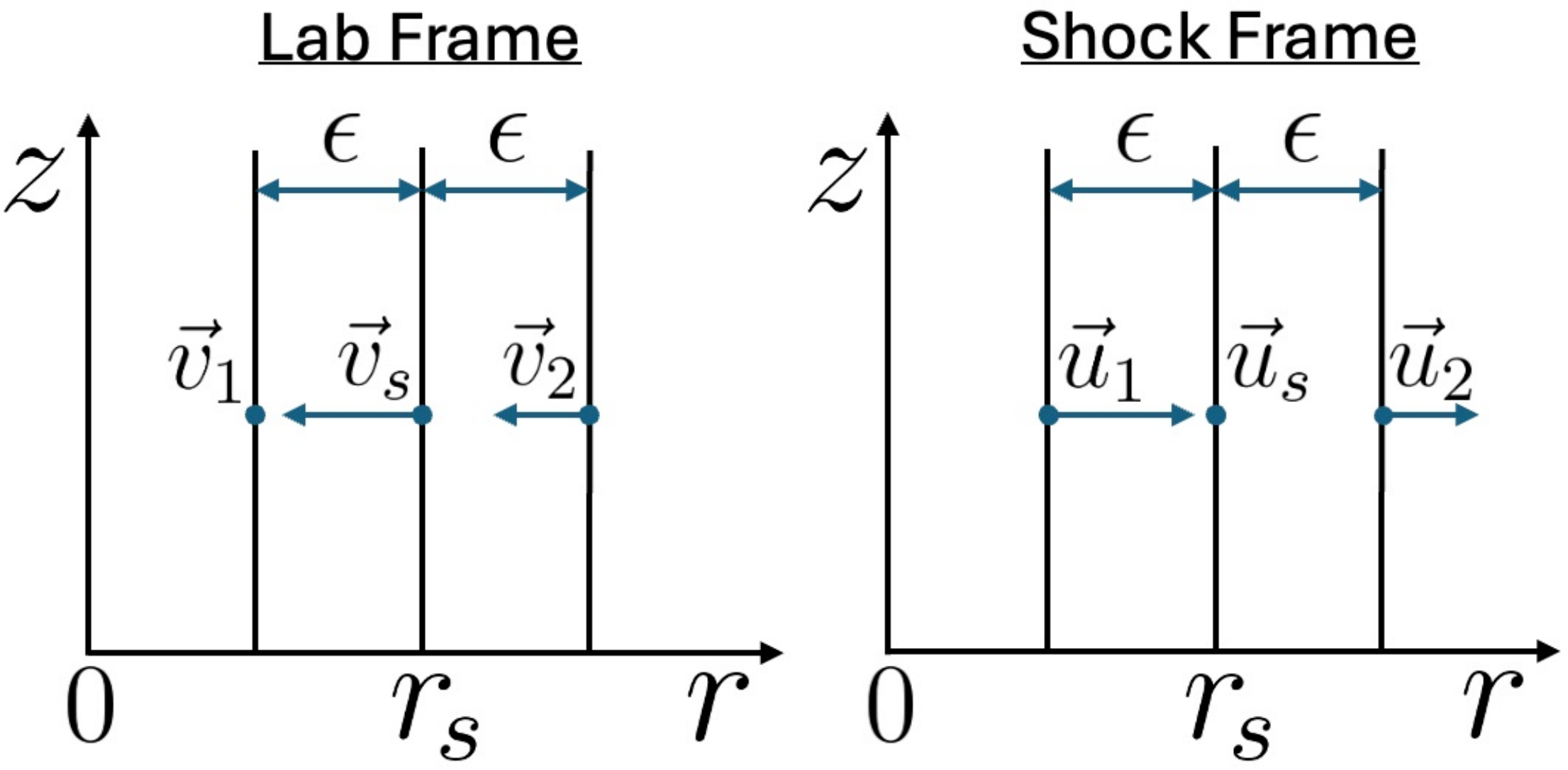}
\caption{Lab Frame vs Shock Frame. The laboratory frame is the standard external frame in which the shock is moving inward with velocity $\vec{v}_{s}$. The shock frame is the comoving frame in which the shock is stationary. We use $\vec{v}$ and $\vec{u}$ to denote flow velocity in the laboratory and shock frames, respectively, with transformation law $\vec{u} = \vec{v}-\vec{v}_{s}$.\label{Fig:frames}}
\end{figure}

During the separation and inward stages, the strong-shock limit $M_{1}^{2}\gg1$ applies due to the relatively small initial pressure $P_{1}$. We also consider the weak initial axial field regime, as is appropriate for most experiments, to take $\beta_{1}\gtrsim 1$. Our ordering is thus $1/M_{1}^{2} \ll 1\lesssim \beta_{1}$. Considering the leading order term to be $\mathcal{O}(1)$, we drop terms at $\mathcal{O}(1/M_{1}^{2})$ and $\mathcal{O}(1/\beta_{1}M_{1}^{2})$. Expanding the solution in this ordering, one finds,
\begin{align}&\label{RH:X}X \sim \frac{\gamma\!+\!1}{\gamma\!-\!1}+\mathcal{O}\!\left(\!\frac{1}{M_{1}^{2}},\frac{1}{\beta_{1}M_{1}^{2}}\!\right),\end{align}
Using Eqs.~(\ref{21:efield},\ref{21:momentum-3}), this corresponds to pressure jump,
\begin{align}&\label{RH:inward-pressure}P_{2}\sim\frac{2}{\gamma\!+\!1}\rho_{1}u_{r1}^{2}+\mathcal{O}\!\left(\!P_{1},\!\frac{P_{1}}{\beta_{1}}\!\right)\!.\!\!\!\!\end{align}
The density, flow velocity, and axial field jumps can be obtained now as well from Eqs.~(\ref{21:mass}-\ref{21:efield}) using Eq.~(\ref{RH:X}),
\begin{align}\label{RH:rho-u-Bz}\frac{\rho_{2}}{\rho_{1}} = \frac{u_{r1}}{u_{r2}} = \frac{B_{z2}}{B_{z1}}\sim\frac{\gamma\!+\!1}{\gamma\!-\!1}+\mathcal{O}\!\left(\!\frac{1}{M_{1}^{2}},\frac{1}{\beta_{1}M_{1}^{2}}\!\right)\!.\end{align}
However, for the flow velocity there is a subtlety related to frame conversion. The RH jump conditions are written in the shock frame, the frame of reference in which the shock is stationary. This is as opposed to the laboratory frame, the standard external frame in which the shock is moving inward with velocity $\vec{v}_{s} \!=\! \dot{r}_{s}\mathbf{\hat{r}}$. The difference is illustrated in Fig.~\ref{Fig:frames}. Then we find,
\begin{align}
&|\vec{v}_{2}-\vec{v}_{s}| = \frac{1}{X}|\,\cancel{\vec{v}_{1}}-\vec{v}_{s}|\rightarrow|\dot{r}_{s}|-|\vec{v}_{2}| = \frac{1}{X}|\dot{r}_{s}|,\end{align}
and taking into account that $\dot{r}_{s} < 0$ and $\mathbf{\hat{r}}\cdot\vec{v}_{2} < 0$,
\begin{align}\label{RH:labv-jump}&\vec{v}_{2} \!=\! \left(\!1-\frac{1}{X}\!\right)\!\dot{r}_{s}\mathbf{\hat{r}} \sim \frac{2}{\gamma\!+\!1}\dot{r}_{s}\mathbf{\hat{r}}+\mathcal{O}\!\left(\!\frac{\dot{r}_{s}}{M_{1}^{2}},\!\frac{\dot{r}_{s}}{\beta_{1} M_{1}^{2}}\!\right),\end{align}
is the true postshock velocity in the laboratory frame. It is worth noting that the ordering used in this section can be expressed as follows,
\begin{align}\mathcal{O}\!\left(\!\frac{1}{M_{1}^{2}},\!\frac{1}{\beta_{1} M_{1}^{2}}\!\right)\sim\mathcal{O}\!\left(\!\frac{v_{S1}^{2}}{\dot{r}_{s}^{2}},\!\frac{v_{A1}^{2}}{\dot{r}_{s}^{2}}\!\right),\end{align}
where $v_{S1}\equiv\sqrt{\gamma P_{1}/\rho_{1}}$ and $v_{A1}\equiv\sqrt{B_{z1}^{2}/\mu_{0}\rho_{1}}$ can be recognized as the upstream sound and Alfv\'en speeds, respectively. This reveals that the strong-shock limit considered here is equivalent to requiring that the shock speed be much greater than both the upstream sound and Alfv\'en speeds, that is, $\dot{r}_{s} \gg v_{S1},v_{A1}$.
\newpage\,\newpage
\section{Appendix B: Velocity Profiles}
\label{App:B}
\renewcommand{\theequation}{B\arabic{equation}}
\renewcommand{\theHequation}{B\arabic{equation}}
\setcounter{equation}{0}
\newcounter{App:B}
We derive in this appendix the approximate velocity profiles throughout different stages of the implosion. In general, these can be obtained in two ways, either directly from the adiabatic law or by differentiating the fluid element trajectories from Appendix~\hyperref[App:D]{D} and interpolating.
\vspace{-10pt}
\subsection{I. Separation \& Inward Stages}
\label{B:I}
We begin with the adiabatic law from Eq.~(\ref{Eq:MHD-3}) in the postshock ($r_{s} < r < r_{p}$) region,
\begin{align}\label{B:adiabatic}\frac{1}{P_{l}(t)}\frac{dP_{l}(t)}{dt}\simeq-\frac{\gamma}{r}\frac{\partial}{\partial r}(rv_{l}(r,t)).\end{align}
Integrating Eq.~(\ref{B:adiabatic}) over $r\in (r_{s},r)$ yields,
\begin{align}\label{B:Pl-rs-r}
\frac{1}{P_{l}(t)}\frac{dP_{l}(t)}{dt}\frac{r^{2}\!-\!r_{s}^{2}}{2\gamma}\simeq r_{s} v_{l}(r_{s}^{+}\!,t)\!-\!r v_{l}(r,t).
\end{align}
Integrating Eq.~(\ref{B:adiabatic}) over $r\in (r,r_{p})$ yields,
\begin{align}\label{B:Pl-r-rp}
\frac{1}{P_{l}(t)}\frac{dP_{l}(t)}{dt}\frac{r_{p}^{2}\!-\!r^{2}}{2\gamma}\simeq r v_{l}(r,t)\!-\!r_{p} v_{l}(r_{p}^{-}\!,t).
\end{align}
Eliminating $(dP_{l}(t)/dt)/P_{l}(t)$ between Eqs.~(\ref{B:Pl-rs-r},\ref{B:Pl-r-rp}) and recognizing $v_{l}(r_{p}^{-}\!,t)=\dot{r}_{p}$ yields the velocity profile,
\begin{align}\label{B:vl-pre}v_{l}(r,t)\simeq\frac{(r^{2}\!-\!r_{s}^{2})r_{p}\dot{r}_{p}\!+\!(r_{p}^{2}\!-\!r^{2})r_{s}v_{l}(r_{s}^{+}\!,t)}{(r_{p}^{2}\!-\!r_{s}^{2})r}.\end{align}
From our RH jump condition in Eq.~(\ref{RH:labv-jump}),
\begin{align}\label{B:RH-v}v_{l}(r_{s}^{+}\!,t)\sim\frac{2}{\gamma\!+\!1}\dot{r}_{s}+\mathcal{O}\!\left(\!\frac{\gamma P_{0}(r_{s})}{\rho_{0}(r_{s})\dot{r}_{s}},\!\frac{B_{z0}^{2}}{\mu_{0}\rho_{0}(r_{s})\dot{r}_{s}}\!\right).\end{align}
Combining Eqs.~(\ref{B:vl-pre},\ref{B:RH-v}) yields,
\begin{align}\label{B:vl}v_{l}(r,t)=\!\frac{(r^{2}\!-\!r_{s}^{2})r_{p}\dot{r}_{p}\!+\!(r_{p}^{2}\!-\!r^{2})\frac{2}{\gamma+1}r_{s}\dot{r}_{s}}{(r_{p}^{2}\!-\!r_{s}^{2})r}.\!\!\!\end{align}
Using the shock ODE in Eq.~(\ref{Eq:Angus-rs}) for this stage, this can also be rewritten,
\begin{align}\label{B:vl-alt}\!\!v_{l}(r,t)\!=\!\frac{2}{\gamma\!+\!1}\frac{r_{s}\dot{r}_{s}}{r}\!-\!\frac{1}{\gamma r}\!\!\left(\!\frac{\ddot{r}_{s}}{\dot{r}_{s}}\!+\!\frac{1}{2}\frac{d}{dt}\ln\rho_{0}(r_{s})\!\!\right)\!\!(r^{2}\!\!-\!r_{s}^{2}),\!\!\!\!\end{align} which generalizes Eq.~(24) in Angus et al.~\cite{Angus}. Alternatively, one could differentiate the fluid element trajectories in either of Eqs.~(\ref{D:ri},\ref{D:ri-alt}) to obtain,
\begin{align}\label{B:v-potter}&\!\!\!\!v_{i}(t_{si} \!< \!t <\! t_{c}) \!\simeq\! \frac{(r_{i}(t)^{2}\!-\!r_{s}^{2})r_{p}\dot{r}_{p}\!+\!(r_{p}^{2}\!-\!r_{i}(t)^{2})\frac{2}{\gamma+1}r_{s}\dot{r}_{s}}{(r_{p}^{2}\!-\!r_{s}^{2})r_{i}(t)}.\end{align}
The profiles in Eqs.~(\ref{B:vl},\ref{B:v-potter}) exactly agree.
\subsection{II. Compression Stage}
\label{B:II}
We begin with the adiabatic law from Eq.~(\ref{Eq:MHD-3}) in the postshock ($0 \leq r < r_{p}$) region,
\begin{align}\label{B:adiabatic-lt}&\!\!\frac{1}{P_{l}(t)}\frac{dP_{l}(t)}{dt} \!\simeq\! -\frac{\gamma}{r}\frac{\partial}{\partial r}(rv_{l}(r,t)).\end{align}
Integrating Eq.~(\ref{B:adiabatic-lt}) over $r\in [0,r)$ yields,
\begin{align}\label{B:Pl-rs-r-lt}
\frac{1}{P_{l}(t)}\frac{dP_{l}(t)}{dt}\frac{r^{2}}{2\gamma}\simeq -r v_{l}(r,t).
\end{align}
Integrating Eq.~(\ref{B:adiabatic-lt}) over $r\in (r,r_{p})$ yields,
\begin{align}\label{B:Pl-r-rp-lt}
\frac{1}{P_{l}(t)}\frac{dP_{l}(t)}{dt}\frac{r_{p}^{2}\!-\!r^{2}}{2\gamma}\simeq r v_{l}(r,t)\!-\!r_{p} v_{l}(r_{p}^{-}\!,t).
\end{align}
Eliminating $(dP_{l}(t)/dt)/P_{l}(t)$ between Eqs.~(\ref{B:Pl-rs-r-lt},\ref{B:Pl-r-rp-lt}) and recognizing $v_{l}(r_{p}^{-}\!,t)=\dot{r}_{p}$ yields the velocity profile,
\begin{align}\label{B:vl-r}v_{l}(r,t)\simeq\frac{r\dot{r}_{p}}{r_{p}}.\end{align}
Alternatively, one could differentiate the fluid element trajectories in Eq.~(\ref{D:rit-compression-1}), which yields,
\begin{align}\label{B:v-potter-compression-1}&v_{i}(t_{c}\! \leq\! t \leq\! t_{p}) \!\simeq\! \frac{r_{i}(t)\dot{r}_{p}}{r_{p}}\simeq\frac{r_{i}(t_{c})\dot{r}_{p}}{r_{p}(t_{c})}.\end{align}
The profiles in Eqs.~(\ref{B:vl-r},\ref{B:v-potter-compression-1}) exactly agree.\\
\indent Thus, in Eqs.~(\ref{B:vl},\ref{B:vl-r}) and Eqs.~(\ref{B:v-potter},\ref{B:v-potter-compression-1}), we have respectively found the Eulerian and Lagrangian velocity profiles in all regions, for all times, in terms of known quantities as outlined in Eq.~(\ref{Eq:model1}).
\subsection*{III. Boundary Layer}
\label{B:III}
\begin{figure}
\centering
\noindent\includegraphics[width=\columnwidth,height=0.66\columnwidth]{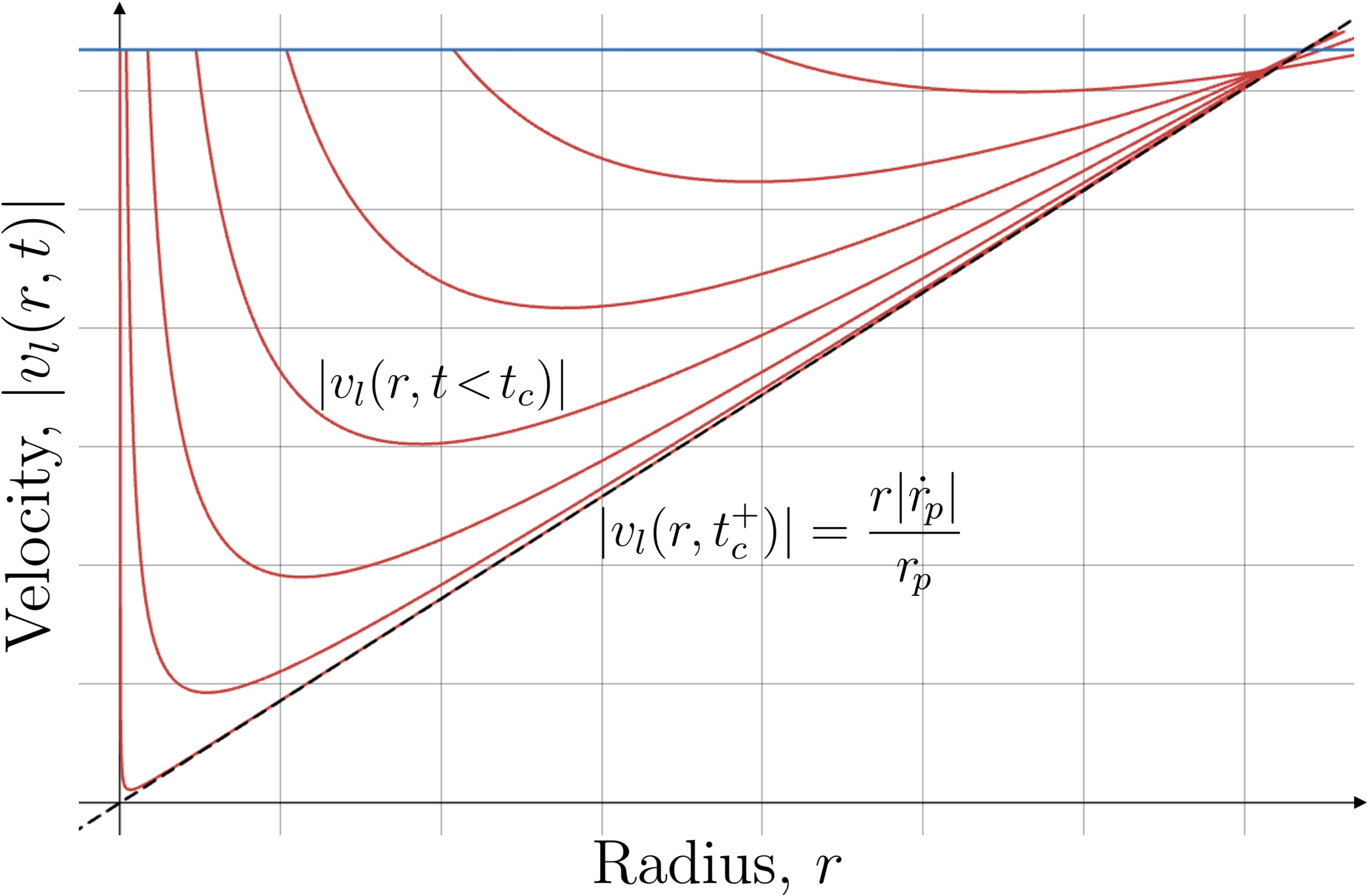}
\caption{Boundary Layer. Shown here is the velocity profile in Eq.~(\ref{B:vl}) for a few moments before $t = t_{c}$ (red) and the velocity profile in Eq.~(\ref{B:vl-r}) immediately afterwards at $t = t_{c}^{+}$ (black). We observe the formation of a boundary layer for $r \gtrsim r_{s}$ as Eq.~(\ref{B:vl}) struggles to fulfill the boundary condition imposed by the RH jump condition in Eq.~(\ref{RH:labv-jump}) at $r = r_{s}^{+}$.\label{Fig:boundary-layer}}
\end{figure}

The transition between the inward and compression stages is quite interesting in that the solution for the velocity profile exhibits boundary layer behavior as $t\rightarrow t_{c}^{-}$ leading to a semicontinuous transition between the velocity profile in the postshock region for the inward stage in Eq.~(\ref{B:vl}) and that for the compression stage in Eq.~(\ref{B:vl-r}). The velocity profile in Eq.~(\ref{B:vl}) has a minimum in magnitude between $r\in (r_{s},r_{p})$ whenever,
\begin{align}(r_{p}^{2}+r_{s}^{2})\dot{r}_{p}\dot{r}_{s}>r_{p}r_{s}\max\left(\!(\gamma+1)\dot{r}_{p}^{2},\frac{4}{\gamma+1}\dot{r}_{s}^{2}\!\right)\!,\end{align}
which evidently always occurs as $r_{s}\rightarrow 0^{+}$. This minimum occurs at $r = \lambda(t)$ where,
\begin{align}\lambda(t) \equiv \sqrt{r_{p}r_{s}}\sqrt{\frac{\frac{2}{\gamma+1}r_{p}\dot{r}_{s}\!-\!r_{s}\dot{r}_{p}}{r_{p}\dot{r}_{p}\!-\!\frac{2}{\gamma+1}r_{s}\dot{r}_{s}}},\end{align}
and corresponds to velocity,
\begin{align}v(\lambda,t)=\frac{2\lambda}{r_{p}^{2}-r_{s}^{2}}\!\left(\!r_{p}\dot{r}_{p}-\frac{2}{\gamma+1}r_{s}\dot{r}_{s}\!\right)\!.\end{align}
Evidently both $\lambda,|v(\lambda,t)|\rightarrow 0^{+}$ as $r_{s}\rightarrow 0^{+}$, but we can see from Eq.~(\ref{B:vl-pre}) that $v(r_{s}^{+}\!,t) \simeq \frac{2}{\gamma+1}\dot{r}_{s}(t_{c}^{-})$ as dictated by the RH jump condition in Eq.~(\ref{RH:labv-jump}). This apparent contradiction corresponds to the formation of a boundary layer~\cite{Bender} near the axis as illustrated in Fig.~\ref{Fig:boundary-layer}.

This boundary layer resolves in the moment of incidence to yield precisely the linear velocity profile in the postshock region during the compression stage in Eq.~(\ref{B:vl-r}). Whether this behavior is observable remains to be seen in comparisons with experimental data, but energy is conserved and the velocity varies continuously at every point in space except on axis for an infinitesimal moment, so it is both physically conceivable and mathematically consistent.\\
\indent Since the velocity profiles in Eqs.~(\ref{B:vl-pre},\ref{B:vl-r}) do not explicitly depend on any of the system parameters and are purely solutions of the adiabatic law for uniform pressure profile, this boundary layer behavior always occurs under the assumptions of our model. It can be viewed as the mathematical manifestation of the instantaneous cylindrical collision that occurs on axis at the moment of incidence.\newpage
\subsection*{IV. Conservation of Energy}
\label{B:IV}
One might wonder if this boundary layer behavior conserves energy in the moment of incidence, and therefore whether one must consider a reflected shock. The total mechanical energy of the plasma in an infinitesimal column of radius $r = \delta$ is given by,
\begin{align}\label{F:Wpt}&\!\!\!\!\frac{U_{\delta}(t)}{2\pi}\!\sim\!\!\!\int\limits_{r_{s}}^{\delta}\!\!\!\left(\!\frac{1}{2}\rho(r,t) v_{r}(r,t)^{2}\!\!+\!\frac{P(r,t)}{\gamma\!-\!1}\!\right)\!rdr\!+\!\mathcal{O}(r_{s}^{2}P_{0}(r)\!).\!\!\!\!\!\end{align}
At the moment of incidence, the pressure profiles in Eqs.~(\ref{C:Pl},\ref{C:Pl-r}) share the value in Eq.~(\ref{C:Pltr}). The density profiles described in Eqs.~(\ref{D:rhoit},\ref{D:rhoil-r}) are also continuous. The possible energy loss can only occur in an infinitesimal column around the axis, so $\delta$ can be made arbitrarily small such that we may consider,
\begin{align}\lim\limits_{\delta\rightarrow 0^{+}}\rho(0\!\leq\! r\! < \!\delta,t_{c}) \simeq \frac{\gamma\!+\!1}{\gamma\!-\!1}\rho_{0}(0),\end{align}
without loss of generality. Then to show that energy is conserved by the discontinuous change in velocity profile, we must merely demonstrate the equality,
\begin{align}\label{B:onaxiscoe}\lim\limits_{r_{s}\rightarrow0^{+}}\int\limits_{r_{s}}^{\delta}\!v_{r}(r,t_{c}^{-})^{2}rdr=\int\limits_{0}^{\delta}\!v_{r}(r,t_{c}^{+})^{2}rdr = \frac{\delta^{4}\dot{r}_{p}^{2}}{4r_{p}^{2}}.\end{align}
which holds for the velocity profiles in Eqs.~(\ref{B:vl},\ref{B:vl-r}). Therefore, no energy remains for the additional jump in pressure associated with a reflected shock wave. This can be understood as a consequence of dynamically imposing a uniform pressure profile, which corresponds to the infinite sound speed limit and thus predicts that all of the inward shock front's energy is carried away by sound waves in the moments preceding its incidence on axis.
\newpage
\section{Appendix C: Pressure Profiles}
\label{App:C}
\renewcommand{\theequation}{C\arabic{equation}}
\renewcommand{\theHequation}{C\arabic{equation}}
\setcounter{equation}{0}
We derive in this appendix the approximate pressure profiles throughout different stages of the implosion. To this end, we begin with the simple principle that in the postshock plasma, the sound speed is sufficiently high that imbalances in pressure quickly resolve themselves to yield an approximately uniform pressure profile.
\subsection{I. Uniform Pressure Approximation}
\label{C:I}
\indent One can justify this assumption of uniform pressure within the postshock region via the following scaling argument. The time it takes for sound waves to cross the sheath $\tau_{Sl}$ scales for downstream sound speed $v_{Sl}$ as,
\begin{align}&\tau_{Sl} \sim \mathcal{O}\!\left(\!\frac{r_{p}\!-\!r_{s}}{v_{Sl}}\!\right) \!\sim \mathcal{O}\!\left(\!(r_{p}\!\!-\!r_{s})\sqrt{\!\frac{\rho(r_{s}^{+})\!\!}{\gamma P(r_{s}^{+})}}\,\right).\end{align}
Using the RH jump conditions in Eqs.~(\ref{RH:inward-pressure},\ref{RH:rho-u-Bz}),
\begin{align}\label{E2:tauS2}&\tau_{Sl} \sim \mathcal{O}\!\left(\!\frac{r_{p}\!\!-\!r_{s}}{|\dot{r}_{s}|}\!\right).\end{align}
The implosion time $\tau_{0}$ roughly scales as,
\begin{align}\label{E2:tau0}&\tau_{0}\sim\mathcal{O}\!\left(\frac{r_{0}}{|\dot{r}_{s}|}\right).\end{align}
Therefore, as long as the sheath remains much narrower than the initial radius $r_{0}$,
\begin{align}\label{E2:tauS2-tau0}&\frac{\tau_{Sl}}{\tau_{0}}\sim\mathcal{O}\!\left(\!\frac{r_{p}\!-\!r_{s}}{r_{0}}\!\right)\rightarrow \tau_{Sl} \ll \tau_{0}.\end{align}
Thus, sound waves should have time to cross the sheath several times throughout the duration of the implosion, equilibrating the pressure profile to a good approximation. Together with the early time reduction of the Angus model for the inward stage to Potter's model for the separation stage, which assumes material pressure equilibration within the sheath, Eq.~(\ref{E2:tauS2-tau0}) provides solid justification for the pressure profile within the postshock region to be approximately uniform, as stated in Eq.~(\ref{Eq:model2}).\\
\indent One could also ask how the sound crossing time compares to the time it takes for the piston radius to change significantly. The natural piston timescale is $\tau_{p}\equiv r_{p}/|\dot{r}_{p}|$, in which case,
\begin{align}\frac{\tau_{Sl}}{\tau_{p}} \sim \mathcal{O}\left(\!\frac{M_{p}}{A_{p}}\!\right),\end{align}
where $M_{p}\equiv |\dot{r}_{p}|/v_{Sl}$ is the piston Mach number with respect to the sheath sound speed, and $A_{p}\equiv r_{p}/(r_{p}-r_{s})$ is the in-flight aspect ratio (IFAR) of the sheath~\cite{Basko,Ruiz}. Requiring $\tau_{Sl}\ll \tau_{p}$ is stronger than $\tau_{Sl}\ll \tau_{0}$, but can be shown to hold as well, particularly during the early and intermediate stages of the implosion, using the dominant balance in Eq.~(\ref{Eq:geometry}).
\subsection*{II. Separation \& Inward Stages}
\label{C:II}
Throughout the separation and inward stages, the uniform pressure profile within the postshock ($r_{s} < r < r_{p}$) region is dictated by the RH jump condition in Eq.~(\ref{RH:inward-pressure}),
\begin{align}\label{C:Pl}P_{l}(t)\sim\frac{2}{\gamma\!+\!1}\rho_{0}(r_{s})\dot{r}_{s}^{2}\!+\!\mathcal{O}\!\left(\!P_{0}(r_{s}),\!\frac{B_{z0}^{2}}{2\mu_{0}}\!\right).\end{align}
The initial pressure profile $P_{0}(r)$ in the unperturbed ($0 \leq r < r_{s}$) region is unknown and can be considered to be determined by the ideal gas law,
\begin{align}\label{C:P0}P_{0}(r)\simeq \frac{\rho_{0}(r)}{m_{i}}k_{B}T_{0}(r),\end{align}
where $m_{i}$ is the ion mass, $k_{B}$ is Boltzmann's constant, and $T_{0}(r)$ is the unknown initial temperature profile. Fortunately, knowledge of $P_{0}(r)$ is not necessary to characterize an implosion in the strong-shock limit, as described in Appendix~\hyperref[A:I]{A.I}.
\subsection*{III. Compression Stage}
\label{C:III}
During the compression stage, the RH jump condition in Eq.~(\ref{RH:inward-pressure}) no longer applies, so we must derive the pressure profile through other means. We begin with the adiabatic law from Eq.~(\ref{Eq:MHD-3}) in the postshock ($0 \leq r < r_{p}$) region,
\begin{align}\label{C:adiabatic-l}\!\!\frac{1}{P_{l}(t)}\frac{dP_{l}(t)}{dt} \!\simeq\! -\frac{\gamma}{r}\frac{\partial}{\partial r}(rv_{l}(r,t)).\!\!\end{align}
Substituting the velocity profile for the compression stage from Eq.~(\ref{B:vl-r}) into Eq.~(\ref{C:adiabatic-l}), we obtain,
\begin{align}\label{C:adiabatic-l2}\!\!\frac{1}{P_{l}(t)}\frac{dP_{l}(t)}{dt} \!\simeq\! -2\gamma\frac{\dot{r}_{p}}{r_{p}}.\end{align}
Integrating Eq.~(\ref{C:adiabatic-l2}) over $t\in(t_{c},t)$ yields,
\begin{align}\label{C:Pl-r}P_{l}(t) \simeq P_{l}(t_{c})\!\!\left(\!\frac{r_{p}(t_{c})}{r_{p}(t)}\!\right)^{\!\!2\gamma},\end{align}
where the RH jump condition in Eq.~(\ref{RH:inward-pressure}) implies that,
\begin{align}\label{C:Pltr}P_{l}(t_{c})\sim\frac{2}{\gamma\!+\!1}\rho_{0}(0)\dot{r}_{s}(t_{c}^{-})^{2}\!+\!\mathcal{O}\!\left(\!P_{0}(0),\!\frac{B_{z0}^{2}}{2\mu_{0}}\!\right).\end{align}
Then Eqs.~(\ref{C:Pl-r},\ref{C:Pltr}) describe the pressure profile in the postshock region in terms of known quantities. The form of Eq.~(\ref{C:Pl-r}) suggests that the piston is compressing the entire column it encloses, adiabatically. That is, Eq.~(\ref{C:Pl-r}) is equivalent to,
\begin{align}P_{l}(t)V_{p}(t)^{\gamma} = P_{l}(t_{c})V_{p}(t_{c})^{\gamma},\end{align}
where $V_{p}(t)\equiv \pi r_{p}^{2}$ is the volume per unit length of this column.
Thus, in Eqs.~(\ref{C:Pl},\ref{C:Pl-r}), we have found the uniform pressure profiles in all regions, for all times, in terms of known quantities, with the exception of the initial pressure profile in Eq.~(\ref{C:P0}), which does not play a significant role in determining the implosion.
\newpage
\section{Appendix D: Lagrangian Profiles}
\label{App:D}
\renewcommand{\theequation}{D\arabic{equation}}
\renewcommand{\theHequation}{D\arabic{equation}}
\setcounter{equation}{0}
\vspace{-5pt}
We derive in this appendix the approximate density and axial field profiles throughout different stages of the implosion, as well as the positions and radial widths of all the fluid elements. We assume that the shock is the only entropic agent in the system, namely that fluid elements on either side of the shock front evolve adiabatically via Eq.~(\ref{Eq:MHD-3}) with the axial field frozen-in according to Alfv\'en's theorem in Eq.~(\ref{Eq:MHD-4}).
\vspace{-10pt}
\subsection*{I. Lagrangian Coordinates}
\vspace{-5pt}
\indent In order to follow individual fluid elements, let us define a 1D Lagrangian coordinate system where the $i$th fluid element has position $r_{i}(t)$, radial width $dr_{i}(t)$, velocity $v_{i}(t)$, pressure $P_{i}(t)$, density $\rho_{i}(t)$, and axial field $B_{zi}(t)$. Its mass $dm_{i}(t)$ and volume $dV_{i}(t)$ per length are,\!
\begin{align}\label{D:dV}dm_{i}(t) = \rho_{i}(t)dV_{i}(t) = 2\pi \rho_{i}(t) r_{i}(t)dr_{i}(t).\end{align}
We define these elements to conserve mass via, \begin{align}\label{D:dm}\frac{dm_{i}(t)}{dt} = 0.\end{align}
Since $\vec{B}\cdot\mathbf{\hat{r}} = 0$, Eqs.~(\ref{Eq:MHD-1},\ref{Eq:MHD-3},\ref{Eq:MHD-4}) can be rewritten,
\begin{align}&\left(\!\frac{\partial}{\partial t}+\vec{v}\cdot\vec{\nabla}\!\right)\!\rho=-\rho\,\vec{\nabla}\cdot\vec{v},\\
&\left(\!\frac{\partial}{\partial t}\!+\!\vec{v}\cdot\vec{\nabla}\!\right)\!P=-\gamma P\,\vec{\nabla}\cdot\vec{v},\\
&\left(\!\frac{\partial}{\partial t}\!+\!\vec{v}\cdot\vec{\nabla}\!\right)\!B_{z} = -B_{z}\,\vec{\nabla}\cdot\vec{v}.\end{align}
Then along characteristic paths $r_{i}(t)$ satisfying,
\begin{align}\label{D:char}\frac{dr_{i}(t)}{dt} = v_{i}(t) = v_{r}(r_{i}(t),t),\end{align}
namely in the Lagrangian frame traveling with the fluid,
\begin{align}\label{D:rho-P-Bz}\!\!\frac{d}{dt}\ln\rho_{i}(t) \!=\! \frac{d}{dt}\ln P_{i}(t)^{1/\gamma} \!=\! \frac{d}{dt}\ln B_{zi}(t).\!\!\!\!\end{align}
This relation holds whenever the ideal MHD equations hold, namely any time except for when a fluid element encounters the shock front. In those moments, we must refer to the RH jump conditions in Appendix~\hyperref[App:A]{A}.
\vspace{-10pt}
\subsection{II. Separation \& Inward Stages}
\label{D:II}
\vspace{-5pt}
We proceed under the assumption that the $i$th fluid element remains untouched prior to the shock reaching its initial position at time $t_{si}$, that is,
\begin{align}
\label{D:Pi0}&P_{i}(0\!\leq\! t \!<\! t_{si})\!=\! P_{i}(0)\!=\! P_{0}(r_{i}(0)),\\
\label{D:rhoi0}&\rho_{i}(0\!\leq\!t\!<\!t_{si}) \!=\! \rho_{i}(0) \!=\! \rho_{0}(r_{i}(0)\!),\\
\label{D:Bzi0}&B_{zi}(0\!\leq\!t \!<\! t_{si})\!=\! B_{z0},\\
\label{D:ri0}&r_{i}(0\!\leq\!t\!\leq\!t_{si}) \!=\! r_{i}(0) \!=\! r_{s}(t_{si}),\\
\label{D:dri0}&dr_{i}(0\!\leq\!t\!<\!t_{si}) \!=\! dr_{i}(0),\\
\label{D:vi0}&v_{i}(0\!\leq\!t\!<\!t_{si}) = 0.\end{align}
At time $t_{si}$ when the element first encounters the shock, the RH jump conditions in Eqs.~(\ref{RH:inward-pressure},\ref{RH:rho-u-Bz}) tell us that,
\begin{align}\label{D:RH}&P_{i}(t_{si}\!<\! t \!<\! t_{c}) = P_{l}(t)\simeq\frac{2}{\gamma\!+\!1}\rho_{0}(r_{s})\dot{r}_{s}^{2},\\
&\frac{\rho_{i}(t_{si}^{+})}{\rho_{i}(0)} = \frac{B_{zi}(t_{si}^{+})}{B_{z0}}\simeq\frac{\gamma\!+\!1}{\gamma\!-\!1}.\!\!\!\end{align}
From Eq.~(\ref{D:rho-P-Bz}), adiabatic evolution and Alfv\'en's theorem in the postshock region yield,
\begin{align}\label{D:rhoit}
\rho_{i}(t_{si}\!<\!t\!<\!t_{c}) = \!\left(\!\frac{P_{i}(t)}{P_{i}(t_{si}^{+})}\!\right)^{\!\!1/\gamma}\!\!\!\!\!\!\!\rho_{i}(t_{si}^{+}),\\
\label{D:Bzit}B_{zi}(t_{si}\!<\!t\!<\!t_{c}) = \frac{\rho_{i}(t)}{\rho_{i}(t_{si}^{+})}B_{zi}(t_{si}^{+}).\end{align}
From Eqs.~(\ref{D:dV},\ref{D:dm}), we obtain the following useful relation between the densities, motions, and widths,
\begin{align}\label{D:motion}r_{i}(t)dr_{i}(t) = \frac{\rho_{i}(0)}{\rho_{i}(t)}r_{i}(0)dr_{i}(0).\end{align}
This is a separable ODE so we can sum, or in the continuous limit integrate over our fluid elements. Let us treat $r_{j}(0)$ as a variable that sweeps across $[0,r_{0}]$, and $r_{j}(t)$ as a variable that ranges across $[0,r_{p}(t)]$ as $j$ ranges over all existing fluid elements. Then Eq.~(\ref{D:motion}) becomes,
\begin{align}\int\limits_{r_{p}(t)}^{r_{i}(t)}\!\!\!\!r_{j}(t)dr_{j}(t) =\!\!\!\!\int\limits_{r_{0}}^{r_{i}(0)}\!\!\!\frac{\rho_{j}(0)}{\rho_{j}(t)}r_{j}(0)dr_{j}(0).
\end{align}
Then, solving for the trajectory of the $i$th fluid element,
\begin{align}\label{D:ri}r_{i}(t_{si}\!<\!t\!<\!t_{c}) = \!\sqrt{r_{p}(t)^{2}-2\!\!\!\!\int\limits_{r_{i}(0)}^{r_{0}}\!\!\frac{\rho_{j}(0)}{\rho_{j}(t)}r_{j}(0)dr_{j}(0)}.\end{align}
This essentially measures where the $i$th fluid element is by adding up the widths of the compressed fluid elements between that element and the piston. Alternatively, we can use the form of the velocity profile in Eq.~(\ref{B:vl}) and directly solve the ODE in Eq.~(\ref{D:char}) to follow a single element and obtain,
\begin{align}\label{D:ri-alt}\!\!\!\!\!r_{i}(t_{si}\!<\!t\!<\!t_{c}) \!=\! \sqrt{r_{p}(t)^{2}\!-\!(r_{p}(t_{si})^{2}\!-\!r_{i}(0)^{2})\!\!\left(\!\frac{P_{l}(t_{si})}{P_{l}(t)}\!\right)^{\!\!1/\gamma}\!}.\!\!\!\!\!\end{align}
The radial width of each fluid element correspondingly evolves as follows,
\begin{align}dr_{i}(t_{si}\!<\!t\!<\!t_{c}) = \frac{r_{i}(0)\rho_{i}(0)}{r_{i}(t)\rho_{i}(t)}dr_{i}(0).\end{align}
The positions in Eqs.~(\ref{D:ri},\ref{D:ri-alt}) are equivalent via Eqs.~(\ref{D:dV},\ref{D:dm}) and can now be used to interpolate the Lagrangian density and axial field profiles in Eqs.~(\ref{D:rhoit},\ref{D:Bzit}) to obtain the Eulerian profiles via $\rho_{l}(r_{i}(t),t) = \rho_{i}(t)$ and $B_{zl}(r_{i}(t),t) = B_{zi}(t)$.

\subsection{III. Compression Stage}
\label{D:III}
\indent During the compression stage, all elements experience the pressure profile in Eq.~(\ref{C:Pl-r}),
\begin{align}\label{D:adiabatic-p}P_{i}(t_{c}\!\leq\!t\!\leq\!t_{p})\simeq P_{l}(t)\simeq P_{l}(t_{c})\!\left(\!\frac{r_{p}(t_{c})}{r_{p}(t)}\!\right)^{\!\!2\gamma}\!\!\!\!\!,\end{align}
which by Eq.~(\ref{D:rho-P-Bz}) yields,
\begin{align}\label{D:rhoil-r}&\rho_{i}(t_{c}\!\leq\!t\!\leq\!t_{p}) \simeq \rho_{i}(t_{c})\frac{r_{p}(t_{c})^{2}}{r_{p}(t)^{2}},\\
\label{D:Bzil-r}&B_{zi}(t_{c}\!\leq\!t\!\leq\!t_{p})\simeq B_{zi}(t_{c})\frac{r_{p}(t_{c})^{2}}{r_{p}(t)^{2}}.\end{align}
Eqs.~(\ref{D:dV},\ref{D:dm}) are once again enough to determine the positions of our fluid elements,
\begin{align}\label{D:motion-l}
&r_{i}(t)dr_{i}(t) \!\simeq\! \frac{r_{p}(t)^{2}}{r_{p}(t_{c})^{2}}r_{i}(t_{c})dr_{i}(t_{c}).\!\!\!\end{align}
Let us treat $r_{j}(t_{c})$ as a variable that sweeps across $[0,r_{p}(t_{c})]$, and $r_{j}(t)$ as a variable that ranges across $[0,r_{p}(t)]$ as $j$ ranges over all existing fluid elements. Then Eq.~(\ref{D:motion-l}) becomes,
\begin{align}&\int\limits_{r_{i}(t)}^{r_{p}(t)}\!\!\!\!r_{j}(t)dr_{j}(t)=\frac{r_{p}(t)^{2}\!}{r_{p}(t_{c})^{2}} \!\!\!\int\limits_{r_{i}(t_{c})}^{r_{p}(t_{c})}\!\!\!\!\!r_{j}(t_{c})dr_{j}(t_{c}),\end{align}
which, in agreement with the direct solution of Eq.~(\ref{D:char}) using the velocity profile in Eq.~(\ref{B:vl-r}), yields,
\begin{align}\label{D:rit-compression-1}r_{i}(t_{c}\!\leq\!t\!\leq\!t_{p}) \simeq \frac{r_{p}(t)}{r_{p}(t_{c})}r_{i}(t_{c}).\end{align}
The radial width of each fluid element correspondingly evolves as follows,
\begin{align}dr_{i}(t_{c}\!\leq\!t\!\leq\!t_{p}) \simeq \frac{r_{i}(t)}{r_{i}(t_{c})}dr_{i}(t_{c}) \simeq \frac{r_{p}(t)}{r_{p}(t_{c})}dr_{i}(t_{c}).\end{align}
The positions in Eq.~(\ref{D:rit-compression-1}) can now be used to interpolate the Lagrangian density and axial field profiles in Eqs.~(\ref{D:rhoil-r},\ref{D:Bzil-r}) to obtain the Eulerian profiles via $\rho_{l}(r_{i}(t),t) = \rho_{i}(t)$ and $B_{zl}(r_{i}(t),t) = B_{zi}(t)$.\\
\indent It may seem curious that the Eulerian velocity and pressure profiles are readily obtainable, but reconstructing the Eulerian density and axial field profiles requires meticulous tracking of individual fluid elements. From a conservation of information standpoint, this must be the case since the density profile within the sheath is a nonuniformly compressed reproduction of the initial density profile $\rho_{0}(r)$, and hence, along with the axial field profile it determines, requires a detailed history of when and how particular fluid elements were encountered by the shock front and subsequently evolved within the sheath.
\newpage\,\newpage
\section{Appendix E: Separation Model}
\label{App:E}
\renewcommand{\theequation}{E\arabic{equation}}
\renewcommand{\theHequation}{E\arabic{equation}}
\setcounter{equation}{0}
We derive in this appendix the first-order coupled ODE system for the separation stage in Eq.~(\ref{Eq:Potter}). This is a generalized and systematized version of the derivation presented in Potter~\cite{Potter}, which assumed constant density and current, in the absence of an axial field.
\vspace{-10pt}
\subsection*{I. Shock ODE}
\label{E:I}
\indent For early times when the sheath is thin and its inertia is negligible, it should be the case that the forces on either side of the sheath approximately equilibrate. Then,
\begin{align}\label{E:uniform}\frac{B_{\theta}^{2}(r_{p}^{+})}{2\mu_{0}} \simeq P(r_{s}^{+}\!,t)\!+\!\frac{B_{z}^{2}(r_{s}^{+}\!,t)}{2\mu_{0}}.\end{align}
The RH jump conditions in Eqs.~(\ref{RH:inward-pressure},\ref{RH:rho-u-Bz}) yield,
\begin{align}\label{E:Ps}\!\!\!\!P(r_{s}^{+}\!,t)\!+\!\frac{B_{z}^{2}(r_{s}^{+}\!,t)}{2\mu_{0}}\!\sim\!\frac{2}{\gamma\!+\!1}\rho_{0}(r_{s})\dot{r}_{s}^{2}\!+\!\mathcal{O}\!\left(\!\!P_{0}(r_{s}),\!\frac{B_{z0}^{2}}{2\mu_{0}}\!\right)\!\!.\!\!\!\!\end{align}
Substituting Eq.~(\ref{E:Ps}) into Eq.~(\ref{E:uniform}) yields,
\begin{align}\label{E:Pp-Ps}\!\!\!\!\!\frac{B_{\theta}^{2}(r_{p}^{+})}{2\mu_{0}}\!\sim\!\frac{2}{\gamma\!+\!1}\rho_{0}(r_{s})\dot{r}_{s}^{2}\!+\!\mathcal{O}\!\left(\!\!P_{0}(r_{s}),\!\frac{B_{z0}^{2}}{2\mu_{0}}\!\right)\!\!.\end{align}
By Amp\`ere's law for an azimuthal loop,
\begin{align}\label{E:ampere}\oint\!\vec{B}\cdot d\vec{l} = \mu_{0}I(t) \rightarrow B_{\theta}(r,t) =\begin{cases}\frac{\mu_{0}I(t)}{2\pi r},&r > r_{p}\\
0, &r < r_{p}\end{cases}.\!\!\!\end{align}
Combining Eqs.~(\ref{E:Pp-Ps},\ref{E:ampere}), solving for $\dot{r}_{s}$, and choosing the $\dot{r}_{s}<0$ branch since the sheath is imploding after all,
\begin{align}\label{E:rsdot}\dot{r}_{s}\!\simeq-\frac{I(t)}{4\pi r_{p}}\sqrt{\!\frac{\mu_{0}(\gamma\!+\!1)}{\rho_{0}(r_{s})}},\end{align}
which is Eq.~(\ref{Eq:Potter}) and corresponds to Eq.~(11) in Potter~\cite{Potter}.
\subsection*{II. Piston ODE}
\label{E:II}
Now let us turn our attention to the evolution of the piston. Recall the MHD adiabatic law from Eq.~(\ref{Eq:MHD-3}),
\begin{align}\label{E:adiabatic-2}\!\!\frac{1}{P_{l}(t)}\frac{dP_{l}(t)}{dt}\!\simeq\!-\frac{\gamma}{r}\frac{\partial}{\partial r}(rv_{l}(r,t)).\!\!\end{align}
Multiplying by $r$ and integrating over $r\in (r_{s},r_{p})$ yields,
\begin{align}\label{E:adiabatic-3}\!\!\frac{1}{P_{l}(t)}\frac{dP_{l}(t)}{dt}\frac{r_{p}^{2}\!-\!r_{s}^{2}}{2\gamma}\!\simeq r_{s}v_{l}(r_{s}^{+}\!,t)\!-\!r_{p}v_{l}(r_{p}^{-}\!,t).\end{align}
Differentiating our expression for $P_{l}(t)$ in Eq.~(\ref{E:Ps}) and again neglecting the axial field terms in the strong-shock ordering, 
\begin{align}\label{E:dP-dt}
\frac{1}{P_{l}(t)}\frac{dP_{l}(t)}{dt} \simeq \frac{2\ddot{r}_{s}}{\dot{r}_{s}}\!+\!\frac{d}{dt}\ln\rho_{0}(r_{s}).\end{align}
Evidently $v_{l}(r_{p}^{-}\!,t)=\dot{r}_{p}$ and the RH jump condition in Eq.~(\ref{RH:labv-jump}) yields,
\begin{align}v_{l}(r_{s}^{+}\!,t)\sim\frac{2}{\gamma\!+\!1}\dot{r}_{s}+\mathcal{O}\!\left(\!\frac{\gamma P_{0}(r_{s})}{\rho_{0}(r_{s})\dot{r}_{s}},\!\frac{B_{z0}^{2}}{\mu_{0}\rho_{0}(r_{s})\dot{r}_{s}}\!\right).\end{align}
Combining Eqs.~(\ref{E:adiabatic-3},\ref{E:dP-dt}) yields,
\begin{align}\label{E:adiabatic-4}\left(\!\frac{\ddot{r}_{s}}{\dot{r}_{s}}\!+\!\frac{1}{2}\frac{d}{dt}\ln\rho_{0}(r_{s})\!\right)\!\frac{r_{p}^{2}\!-\!r_{s}^{2}}{\gamma}\simeq \frac{2}{\gamma\!+\!1}r_{s}\dot{r}_{s}\!-\!r_{p}\dot{r}_{p}.\end{align}
Differentiating Eq.~(\ref{E:rsdot}) with respect to time yields,
\begin{align}\label{E:rsddot}\frac{\ddot{r}_{s}}{\dot{r}_{s}} \simeq \frac{\dot{I}}{I}-\frac{\dot{r}_{p}}{r_{p}}-\frac{1}{2}\frac{d}{dt}\ln\rho_{0}(r_{s}).\end{align}
Combining Eqs.~(\ref{E:adiabatic-4},\ref{E:rsddot}) yields,
\begin{align}\label{E:adiabatic-5}\left(\!\frac{\dot{I}}{I}-\frac{\dot{r}_{p}}{r_{p}}\!\right)\!\frac{r_{p}^{2}\!-\!r_{s}^{2}}{\gamma}\simeq \frac{2}{\gamma\!+\!1}r_{s}\dot{r}_{s}\!-\!r_{p}\dot{r}_{p}.\end{align}
Solving for $\dot{r}_{p}$, we obtain,
\begin{align}\label{E:rpdot}\!\dot{r}_{p}\simeq\frac{\frac{2}{\gamma+1}r_{s}\dot{r}_{s}\!-\!\frac{r_{p}^{2}-r_{s}^{2}}{\gamma}\frac{\dot{I}}{I}}{r_{p}\!-\!\frac{r_{p}^{2}-r_{s}^{2}}{\gamma r_{p}}},\!\!\!\end{align}
which is Eq.~(\ref{Eq:Potter}) and corresponds to Eq.~(10) in Potter~\cite{Potter}. The key departures from Potter's original results are the time-dependent current and initial density specifically evaluated at the shock front in Eq.~(\ref{E:rsdot}), as well as the $\dot{I}/I$ term in Eq.~(\ref{E:rpdot}).\\
\indent Potter's model breaks down as the sheath inertia becomes significant and the inward and outward forces fall out of balance, producing the inward acceleration of the piston. Therefore, it should not be used in general to model the full implosion, and we transition to Eqs.~(\ref{Eq:simple-rp},\ref{Eq:Angus-rs}) as soon as stability allows it.\\
\indent Nevertheless, Potter's model still holds up surprisingly well throughout the implosion for low initial density $\rho_{0}(r)$. In the case of constant initial density and current, and no axial field, it can be integrated to obtain an exact closed-form solution.
\vspace{-10pt}
\subsection{III. Constant Density \& Current}
\vspace{-2pt}
For constant initial density $\rho_{0}(r) = \rho_{0}$ and current $I(t) = I_{0}$, Eq.~(\ref{E:rpdot}) becomes,
\begin{align}\frac{\gamma}{\gamma\!+\!1}\frac{dr_{s}^{2}}{dr_{p}}r_{p} = (\gamma\!-\!1)r_{p}^{2}+r_{s}^{2},\end{align}
which for the initial conditions in Eq.~(\ref{Eq:IC}) yields,
\begin{align}\label{E:rp(rs)}r_{p} = r_{0}\!\left(\frac{\gamma}{\gamma\!+\!1\!-\!r_{s}^{2}/r_{p}^{2}}\right)^{\!\!\frac{\gamma}{\gamma-1}}\!\!\!\!\!\!\rightarrow r_{\text{min}}\simeq r_{0}\!\left(\frac{\gamma}{\gamma\!+\!1}\right)^{\!\!\frac{\gamma}{\gamma-1}}\!,\!\end{align}
which is Eqs.~(13,14) in Potter~\cite{Potter}. Potter's model and Eqs.~(\ref{E:rsdot},\ref{E:rpdot},\ref{E:rp(rs)}) do not hold in general, but the predicted stagnation radius $r_{\text{min}}$ agrees quite well with Figs.~\ref{Fig:5532-r},\ref{Fig:4959-r},\ref{Fig:Bz0-0-r},\ref{Fig:Bz0-0.2-r} despite the rapidly varying current and initial density, and even in the presence of an axial field. It may be used with caution for quick estimates.

\section{Appendix F: Inward Model}
\label{App:F}
\renewcommand{\theequation}{F\arabic{equation}}
\renewcommand{\theHequation}{F\arabic{equation}}
\setcounter{equation}{0}
\newcounter{App:F}
We derive in this appendix the second-order coupled ODE system for the inward stage in Eqs.~(\ref{Eq:simple-rp},\ref{Eq:Angus-rs}) as well as the conservation of energy criterion in Eq.~(\ref{Eq:Angus-Wm}). This is a generalized and systematized version of the derivation presented in Angus et al.~\cite{Angus}, which assumed constant density and current, in the absence of an axial field. We also demonstrate reduction to the first-order Potter's model for the separation stage in Eq.~(\ref{Eq:Potter}), in the singular limit of early times.
\vspace{-12.5pt}
\subsection*{I. Piston ODE}
\vspace{-5pt}
\indent We begin with the ideal MHD momentum equation from Eq.~(\ref{Eq:MHD-2}),
\begin{align}
&\label{F:momentum}\rho\!\left(\frac{\partial}{\partial t}+\vec{v}\cdot\vec{\nabla}\right)\!\vec{v} = \frac{(\vec{\nabla}\times\vec{B})\times\vec{B}}{\mu_{0}}-\vec{\nabla}P.\end{align}
In cylindrical coordinates,
\begin{align}\label{F:curl-B}\vec{\nabla}\times\vec{B} = -\frac{\partial B_{z}}{\partial r}\boldsymbol{\hat{\theta}}+\frac{1}{r}\frac{\partial(rB_{\theta})}{\partial r}\mathbf{\hat{z}}.\end{align}
Then taking $\mathbf{\hat{r}}\,\cdot\,$Eq.~(\ref{F:momentum}) and using Eq.~(\ref{F:curl-B}) yields,
\begin{align}\label{F:momentum3}
&\rho\frac{dv_{r}}{dt} \!=\!-\frac{1}{\mu_{0}}\!\!\left(\!B_{z}\frac{\partial B_{z}}{\partial r}\!+\!\frac{B_{\theta}}{r}\frac{\partial(rB_{\theta})}{\partial r}\!\right)-\frac{\partial P}{\partial r}.\end{align}
Rearranging,
\begin{align}\label{F:2nd-law}
&\rho\frac{dv_{r}}{dt} \!+\!\frac{\partial}{\partial r}\!\!\left(\!P\!+\!\frac{B_{\theta}^{2}\!+\!B_{z}^{2}}{2\mu_{0}}\!\right)\!+\!\frac{B_{\theta}^{2}}{\mu_{0}r}=0.\end{align}
By Amp\`ere's law for an azimuthal loop,
\begin{align}\label{F:ampere}\oint\!\vec{B}\cdot d\vec{l} = \mu_{0}I(t) \rightarrow B_{\theta}(r,t) \!=\!\begin{cases}\frac{\mu_{0}I(t)}{2\pi r},&r > r_{p}\\
0, &r < r_{p}\end{cases}.\end{align}
Then multiplying by $r^{2}$ and integrating Eq.~(\ref{F:2nd-law}) over $r\in(r_{s},r_{p}]$, we obtain,
\begin{align}\label{F:start}\!\int\limits_{r_{s}}^{r_{p}}\!\!r\frac{dv_{r}}{dt}\rho rdr\!+\!\!\!\int\limits_{r_{s}}^{r_{p}}\!\!r^{2}\frac{\partial}{\partial r}\!\!\left(\!P\!+\!\frac{B_{\theta}^{2}\!+\!B_{z}^{2}}{2\mu_{0}}\!\right)\!dr\!=0.\end{align}
We must be careful to include $r = r_{p}$, where the density formally diverges, in order to capture all of the mass, as well as the external azimuthal field force at $r = r_{p}^{+}$ which does not necessarily equal the net material and magnetic pressure at $r = r_{p}^{-}$.
Let us define the mass average,
\begin{align}\label{F:mass-avg}\langle f\rangle_{m} \equiv \frac{1}{m(t)}\!\int \!f dm = \frac{2\pi}{m(t)} \!\!\int\limits_{r_{s}}^{r_{p}}\!f \rho\,r dr,\end{align}
where the mass of the sheath is given by,
\begin{align}\label{F:sheath-mass}m(t)\equiv\!\!\int\limits_{0}^{2\pi}\!\int\limits_{r_{s}}^{r_{p}}\!\rho dV = 2\pi\!\!\int\limits_{r_{s}}^{r_{p}}\!\rho_{l}(r,t) r dr = 2\pi\!\!\int\limits_{r_{s}}^{r_{0}}\!\rho_{0}(r) r dr,\end{align}
and in the second equality we have used conservation of mass as implied by Eq.~(\ref{Eq:MHD-1}). Using Eq.~(\ref{F:mass-avg}) and integrating by parts in Eq.~(\ref{F:start}), we obtain,
\begin{align}\label{F:snowplow-start}\!\!\!\!\frac{m(t)}{2\pi}\!\left\langle\!r\frac{dv_{r}}{dt}\!\right\rangle_{\!\!m}\!\!\!\!+\!\!\left[r^{2}\!\!\left(\!P\!+\!\frac{B_{\theta}^{2}\!+\!B_{z}^{2}}{2\mu_{0}}\!\right)\right|_{r_{s}^{+}}^{r_{p}^{+}}\!\!\!\!\!=2\!\!\int\limits_{r_{s}}^{r_{p}}\!\!\left(\!P\!+\!\frac{B_{z}^{2}\!-\!B_{\theta}^{2}}{2\mu_{0}}\!\right)\!rdr.\!\!\!\end{align}
Since the sheath mass becomes overwhelmingly concentrated near the piston, and the extra factor of $r$ only serves to exaggerate this effect, $\langle r\dot{v}_{r}\rangle_{m}\simeq r_{p}\ddot{r}_{p}$~\cite{Angus}. This aligns with the infinite sound speed approximation in that the full sheath inertia is felt by the piston. Then,
\begin{align}\label{F:pre-ddot-rp}&r_{p}^{2}\frac{B_{\theta}^{2}(r_{p}^{+})}{2\mu_{0}}-\!r_{s}^{2}\!\!\left(\!P(r_{s}^{+})\!+\!\frac{B_{z}^{2}(r_{s}^{+})}{2\mu_{0}}\right)\nonumber\\
&+\!\frac{m(t)}{2\pi}r_{p}\ddot{r}_{p}\!=2\!\!\int\limits_{r_{s}}^{r_{p}}\!\!\left(\!P\!+\!\frac{B_{z}^{2}}{2\mu_{0}}\!\right)\!rdr.\end{align}
The RH jump conditions in Eqs.~(\ref{RH:inward-pressure},\ref{RH:rho-u-Bz}) yield,
\begin{align}\label{F:Ps}P(r_{s}^{+})\!+\!\frac{B_{z}^{2}(r_{s}^{+})}{2\mu_{0}}\!\sim\!\frac{2}{\gamma\!+\!1}\rho_{0}(r_{s})\dot{r}_{s}^{2}\!+\!\mathcal{O}\!\left(\!P_{0}(r_{s}),\!\frac{B_{z0}^{2}}{2\mu_{0}}\!\right)\!\!.\!\!\!\!\end{align}
Then substituting Eq.~(\ref{F:Ps}) into Eq.~(\ref{F:pre-ddot-rp}) yields,
\begin{align}\label{F:pre-ddot-rp2}&r_{p}^{2}\frac{B_{\theta}^{2}(r_{p}^{+})}{2\mu_{0}}\!-\!r_{s}^{2}\frac{2}{\gamma\!+\!1}\rho_{0}(r_{s})\dot{r}_{s}^{2}\!+\!\mathcal{O}\!\left(\!r_{s}^{2}P_{0}(r_{s}),r_{s}^{2}\frac{B_{z0}^{2}}{2\mu_{0}}\!\right)\nonumber\\
&+\frac{m(t)}{2\pi}r_{p}\ddot{r}_{p}\!=2\!\!\int\limits_{r_{s}}^{r_{p}}\!\!\left(\!P\!+\!\frac{B_{z}^{2}}{2\mu_{0}}\!\right)\!rdr.\end{align}
Recalling our azimuthal field from Eq.~(\ref{F:ampere}),
\begin{align}\label{F:pre-ddot-rp3}&\frac{m(t)}{2\pi}r_{p}\ddot{r}_{p}\!\simeq\!-\frac{\mu_{0}I^{2}}{8\pi^{2}}\!+\!\frac{2}{\gamma\!+\!1}\rho_{0}(r_{s})\dot{r}_{s}^{2}r_{s}^{2}\!+\!2\!\!\int\limits_{r_{s}}^{r_{p}}\!\!\left(\!P\!+\!\frac{B_{z}^{2}}{2\mu_{0}}\!\right)\!rdr.\end{align}
Then our only unknown in Eq.~(\ref{F:pre-ddot-rp3}) is the material and magnetic pressure integral. The simplest way to evaluate this quantity is to assume the material pressure is uniform within the sheath in line with Eq.~(\ref{Eq:model2}). Together with Eq.~(\ref{F:Ps}), this yields,
\begin{align}\label{F:simple-rp}&\frac{m(t)}{2\pi}r_{p}\ddot{r}_{p}\!\simeq\!-\frac{\mu_{0}I^{2}}{8\pi^{2}}\!+\!\frac{2}{\gamma\!+\!1}\rho_{0}(r_{s})\dot{r}_{s}^{2}r_{p}^{2}\!+\!2\!\!\int\limits_{r_{s}}^{r_{p}}\!\frac{B_{z}^{2}}{2\mu_{0}}rdr,\!\!\end{align}
which is precisely Eq.~(\ref{Eq:simple-rp}). An alternative method to evaluate this integral employs conservation of energy.
\subsection*{II. Conservation of Energy}
\label{F:II}
The difference between the present sum of the total mechanical energy of the plasma, $U(t)$, and the potential energy stored in the axial field, $\Phi(t)$, and their initial sum, is precisely the work done by the azimuthal field, $W(t)$.
\begin{align}\label{F:COE}U(t) + \Phi(t) = U(0) + \Phi(0)+W(t).\end{align}
We may compute the work done on the plasma via,
\begin{align}\label{F:Wm}&W(t)=2\pi\!\!\!\!\int\limits_{r_{p}(t)}^{r_{0}}\!\!\frac{B_{\theta}^{2}(r_{p}^{+},t)}{2\mu_{0}}r_{p}dr_{p}\!= \!-\frac{\mu_{0}}{4\pi}\!\!\int\limits_{0}^{t}\!\frac{I^{2}\dot{r}_{p}}{r_{p}}dt.\end{align}
Integrating by parts, we finally obtain,
\begin{align}\label{F:xi-B}W(t) =\!\frac{\mu_{0}}{4\pi}\!\left(\Delta(t)\!+\!2\zeta(t)\right),\end{align}
where we have defined,
\begin{align}
&\label{F:delta}\Delta(t)\equiv I(0)^{2}\ln r_{0}\!-\!I(t)^{2}\ln r_{p},\\
&\label{F:zeta}\zeta(t)\equiv\!\!\int\limits_{0}^{t}\!I\dot{I}\ln r_{p}\,dt.
\end{align}
By equipartition theorem and the ideal gas law, the initial total mechanical energy of the plasma is given by,
\begin{align}\label{F:Wp0}U(0)=\frac{f}{2}\!\int\limits_{V}\!\frac{\rho_{0}(\vec{r})}{m_{i}}k_{B}T_{0}(\vec{r})dV \!=\! \frac{2\pi}{\gamma\!-\!1}\!\int\limits_{0}^{r_{0}}\!P_{0}(r)rdr,\!\!\end{align}
where $f \equiv 2/(\gamma-1)$ is the number of degrees of freedom of our plasma, $m_{i}$ is the ion mass, and $T_{0}(\vec{r})$ is the initial temperature distribution. The total mechanical energy of the plasma at any given time is,
\begin{align}\label{F:Wpt}&\!\!\!\!\frac{U(t)}{2\pi}\!=\!\!\!\int\limits_{r_{s}}^{r_{p}}\!\!\!\left(\!\frac{1}{2}\rho(r,t) v_{r}(r,t)^{2}\!\!+\!\frac{P(r,t)}{\gamma\!-\!1}\!\right)\!rdr\!+\!\!\!\int\limits_{0}^{r_{s}}\!\!\frac{P_{0}(r)}{\gamma\!-\! 1}rdr.\!\!\!\!\!\end{align}
We assume that the total mechanical energy of the sheath is split approximately equally between the directed flow energy and the thermal energy associated with the material pressure. This is exactly true for planar geometry and is a good approximation for strong shocks in cylindrical geometry when the thickness of the sheath is small compared to its radius of curvature~\cite{Angus,Allen}.
\begin{align}\label{F:equipartition}
&\!\!U(t)\!\simeq\frac{4\pi}{\gamma\!-\!1}\!\int\limits_{r_{s}}^{r_{p}}\!\!P(r,t)rdr\!+\!\frac{2\pi}{\gamma\!-\! 1}\!\int\limits_{0}^{r_{s}}\!\!P_{0}(r)rdr.\!\!\!\end{align}
The initial energy stored in the axial field is given by,
\begin{align}\label{F:Ub0}&\Phi(0) = 2\pi\!\!\int\limits_{0}^{r_{0}}\frac{B_{z0}^{2}}{2\mu_{0}}rdr = \pi r_{0}^{2}\frac{B_{z0}^{2}}{2\mu_{0}}.\end{align}
The axial field energy at some later time is given by,
\begin{align}\label{F:Ubt}&\Phi(t) \!=\! 2\pi\!\!\int\limits_{r_{s}}^{r_{p}}\!\frac{B_{zl}^{2}(r,t)}{2\mu_{0}}rdr\!+\!\pi r_{s}^{2}\frac{B_{z0}^{2}}{2\mu_{0}}.\end{align}
Then, substituting Eqs.~(\ref{F:Wm}-\ref{F:Ubt}) into Eq.~(\ref{F:COE}) and neglecting subdominant terms, we discover the relation,
\begin{align}\label{F:energy}&\int\limits_{r_{s}}^{r_{p}}\left(\!\frac{2P(r,t)}{\gamma\!-\!1}\!+\!\frac{B_{zl}^{2}(r,t)}{2\mu_{0}}\!\right)rdr\simeq\!\!\int\limits_{r_{p}}^{r_{0}}\!\frac{B_{\theta}^{2}(r_{p}^{+},t)}{2\mu_{0}}r_{p}dr_{p}.\!\end{align}
This statement says that in the strong-shock limit, the total directed flow, thermal, and magnetic energy of the sheath is roughly equal to the work done by the axial current via the azimuthal field. Solving this equation for the unknown pressure integral in Eq.~(\ref{F:pre-ddot-rp3}) yields,
\begin{align}\label{F:energy-soln}&2\!\int\limits_{r_{s}}^{r_{p}}\!\!\left(\!P(r,t)\!+\!\frac{B_{zl}^{2}(r,t)}{2\mu_{0}}\!\right)\!rdr \!=\! \frac{\mu_{0}(\gamma\!-\!1)}{8\pi^{2}}(\Delta(t)\!+\!2\zeta(t))\!\!\nonumber\\
&+\!(3\!-\!\gamma)\!\int\limits_{r_{s}}^{r_{p}}\!\frac{B_{zl}^{2}(r,t)}{2\mu_{0}}rdr\!+\!\mathcal{O}\!\left(\!r_{0}^{2}P_{0}(r),r_{0}^{2}\frac{B_{z0}^{2}}{2\mu_{0}}\right).\end{align}
Using Eq.~(\ref{F:energy-soln}) to evaluate the integral in Eq.~(\ref{F:pre-ddot-rp3}) yields the alternative piston ODE,
\begin{align}\label{F:angus-rp}
&\frac{m(t)}{2\pi}r_{p}\ddot{r}_{p}\simeq -\frac{\mu_{0}I^{2}}{8\pi^{2}}\!+\!\frac{\mu_{0}(\gamma\!-\!1)}{8\pi^{2}}(\Delta(t)\!+\!2\zeta(t))\nonumber\\
&+\!\frac{2}{\gamma\!+\!1}\rho_{0}(r_{s})\dot{r}_{s}^{2}r_{s}^{2}+\!(3\!-\!\gamma)\!\int\limits_{r_{s}}^{r_{p}}\!\frac{B_{zl}^{2}(r,t)}{2\mu_{0}}rdr.\end{align}
The conservation of energy criterion in Eq.~(\ref{F:energy}) can also be used to eliminate the explicit axial field contributions entirely from Eq.~(\ref{F:pre-ddot-rp3}) to yield,
\begin{align}\label{F:pre-novel}
&\frac{m(t)}{2\pi}r_{p}\ddot{r}_{p}\simeq-\frac{\mu_{0}I^{2}}{8\pi^{2}}\!+\!\frac{\mu_{0}}{4\pi^{2}}\!\left(\Delta(t)\!+\!2\zeta(t)\right)\nonumber\\
&+\!\frac{2}{\gamma^{2}\!-\!1}\rho_{0}(r_{s})\dot{r}_{s}^{2}\!\left((\gamma\!-\!3)r_{p}^{2}\!+\!2r_{s}^{2}\right).\end{align}
Thus in Eqs.~(\ref{F:simple-rp},\ref{F:angus-rp},\ref{F:pre-novel}), we have three different versions of the piston ODE which appear entirely interchangeable. They all perform similarly for realistic scenarios, but each has distinct advantages. Eq.~(\ref{F:angus-rp}) is the true generalization of Angus et al.~\cite{Angus} Eq.~(33) and does not require a uniform pressure profile. Eq.~(\ref{F:pre-novel}) does not explicitly require knowledge of the axial field or tracking of the individual fluid elements at all. However, they both require the more nebulous equipartition assumption preceding Eq.~(\ref{F:equipartition}) and the resulting conservation of energy criterion in Eq.~(\ref{F:energy}). We will see in Appendix~\hyperref[F:III]{F.III} that obtaining the shock ODE in Eq.~(\ref{Eq:Angus-rs}) already requires assuming uniform pressure, and we need to track the fluid elements anyway to determine the density and axial field profiles, so using the simpler Eq.~(\ref{F:simple-rp}) as stated in Eq.~(\ref{Eq:simple-rp}) is generally preferable. Eq.~(\ref{F:simple-rp}) is also the most compatible with Eq.~(\ref{Eq:Potter}) for the separation stage, as demonstrated in Appendix~\hyperref[F:IV]{F.IV}, and Eq.~(\ref{Eq:Adiabatic-rp}) for the compression stage.
\subsection*{III. Shock ODE}
\label{F:III}
\indent Now that we have an equation describing evolution of the piston, all that remains is to describe the evolution of the shock. Starting with the adiabatic law in Eq.~(\ref{Eq:MHD-3}),
\begin{align}\label{F:div-v-Ps}\!\!\frac{1}{P_{l}(t)}\frac{dP_{l}(t)}{dt} \!\simeq\! -\frac{\gamma}{r}\frac{\partial}{\partial r}(rv_{l}(r,t)),\!\!\end{align}
where this uniform pressure is once again given by the RH jump condition in Eq.~(\ref{RH:inward-pressure}),
\begin{align}P_{l}(t)\sim\frac{2}{\gamma\!+\!1}\rho_{0}(r_{s})\dot{r}_{s}^{2}\!+\!\mathcal{O}\!\left(\!P_{0}(r_{s}),\frac{B_{z0}^{2}}{2\mu_{0}}\right).\end{align}
In this approximation, we have,
\begin{align}\label{F:dPs/dt}\frac{1}{P_{l}(t)}\frac{dP_{l}(t)}{dt} \simeq\frac{2\ddot{r}_{s}}{\dot{r}_{s}}\!+\!\frac{d}{dt}\ln\rho_{0}(r_{s}).\end{align}
Combining Eqs.~(\ref{F:div-v-Ps},\ref{F:dPs/dt}) with our velocity profile in Eq.~(\ref{B:vl}) and solving for $\ddot{r}_{s}$ yields,
\begin{align}
&\label{F:ddot-rs}\ddot{r}_{s}\simeq-\frac{\gamma\dot{r}_{s}}{r_{p}^{2}\!-\!r_{s}^{2}}\!\left(\!r_{p}\dot{r}_{p}\!-\!\frac{2}{\gamma\!+\!1}r_{s}\dot{r}_{s}\!\!\right)\!-\!\frac{\dot{r}_{s}}{2}\frac{d}{dt}\!\ln\rho_{0}(r_{s}).
\end{align}
\subsection*{IV. Early Times}
\label{F:IV}
We demonstrate here that although the second-order ODE system defined by the generalized Angus model has a singular point at the initial conditions, all three versions of the equations reduce down to the first-order ODE system defined by Potter's model in the limit of early times.\\
\indent Looking at Eqs.~(\ref{F:simple-rp},\ref{F:angus-rp},\ref{F:pre-novel}) for $t\gtrsim 0$ and $r_{s}\lesssim r_{p}\lesssim r_{0}$, let us identify what happens to each of the terms. The inertial mass of the sheath $m(t)$ vanishes as can be seen from its integral definition in Eq.~(\ref{F:sheath-mass}) since its bounds are nearly equal. The axial field integral's bounds are nearly equal as well and $B_{zl}(r,t)\sim\mathcal{O}(B_{z0})$ for early times, making it negligible in the strong-shock limit described in Appendix~\hyperref[A:I]{A.I}. The $\Delta(t), \zeta(t)$ terms defined in Eqs.~(\ref{F:delta},\ref{F:zeta}) are also clearly negligible for early times as can be seen from the expression for their sum in Eq.~(\ref{F:Wm}) using $I(0)\equiv 0$. Our second-order piston ODE in Eq.~(\ref{F:simple-rp}) then approaches the first-order ODE,
\begin{align}\label{F:early-times}
&\frac{\mu_{0}I^{2}}{8\pi^{2}}\!\sim\!\frac{2}{\gamma\!+\!1}\rho_{0}(r_{s})\dot{r}_{s}^{2}r_{p}^{2}+\mathcal{O}\!\left(\!P_{0}(r_{s}),\frac{B_{z0}^{2}}{2\mu_{0}}\right).\end{align}
This is clearly equivalent to the uniform pressure relation in Eq.~(\ref{E:Pp-Ps}) from which the first-order shock ODE of Potter's model in Eq.~(\ref{E:rsdot}) was derived. Eqs.~(\ref{F:angus-rp},\ref{F:pre-novel}) approach Eq.~(\ref{F:early-times}) as well with $r_{s}^{2}$ in place of $r_{p}^{2}$ on the right-hand side, which presents a negligible difference for early times when $r_{s}\simeq r_{p}$.\\
\indent By differentiating Eq.~(\ref{F:early-times}), we can see as in Potter's model Eq.~(\ref{E:rsddot}) that,
\begin{align}\label{F:rsddot}\frac{\ddot{r}_{s}}{\dot{r}_{s}} \simeq \frac{\dot{I}}{I}-\frac{\dot{r}_{p}}{r_{p}}-\frac{1}{2}\frac{d}{dt}\ln\rho_{0}(r_{s}).\end{align}

Substituting Eq.~(\ref{F:rsddot}) into Eq.~(\ref{F:ddot-rs}) and solving for $\dot{r}_{p}$, we see that it reduces down to its Potter's model equivalent as well,
\begin{align}\label{F:rpdot}\!\dot{r}_{p}\simeq\frac{\frac{2}{\gamma+1}r_{s}\dot{r}_{s}\!-\!\frac{r_{p}^{2}-r_{s}^{2}}{\gamma}\frac{\dot{I}}{I}}{r_{p}\!-\!\frac{r_{p}^{2}-r_{s}^{2}}{\gamma r_{p}}}.\!\!\!\end{align}
\indent Evidently then, the initial conditions represent a singular point of the Angus model corresponding to reduction of order and swapping of variables of the governing set of piston and shock radius ODEs. In practice, this means the second-order integration scheme used for Angus' model during the inward stage is unstable for early times and cannot be used to initialize the implosion. This necessitates the introduction of the separation stage governed by Potter's model, as described in the main text.
\newpage
\section{Appendix G: Compression Model}
\label{App:G}
\renewcommand{\theequation}{G\arabic{equation}}
\renewcommand{\theHequation}{G\arabic{equation}}
\setcounter{equation}{0}
\newcounter{App:G}
We derive in this appendix the second-order piston ODE for the compression stage in Eq.~(\ref{Eq:Adiabatic-rp}). This is a novel derivation inspired by the Angus model which, as for the inward stage, employs the momentum and adiabatic MHD equations to describe the implosion.\\
\indent As shown in Appendix~\hyperref[C:III]{C.III}, the pressure in the postshock ($r_{s} < r < r_{p}$) region is given by Eqs.~(\ref{C:Pl-r},\ref{C:Pltr}),
\begin{align}\label{G:Pl-r}P_{l}(t) \!\simeq\! \frac{2}{\gamma\!+\!1}\rho_{0}(0)\dot{r}_{s}(t_{c}^{-})^{2}\!\!\left(\!\frac{r_{p}(t_{c})}{r_{p}(t)}\!\right)^{\!\!2\gamma}.\!\!\!\end{align}
\subsection*{I. Piston ODE}
From the perspective of the piston and the postshock ($0 \leq r < r_{p}$) region, little has changed from the inward stage and the derivation presented in Appendix ~\hyperref[App:F]{F} Eqs.~(\ref{F:momentum}-\ref{F:pre-ddot-rp}) still holds except that now $r_{s} \equiv 0$, leaving us with,
\begin{align}\label{G:pre-ddot-rp}&\frac{\mu_{0}I^{2}}{8\pi^{2}}\!+\!\frac{m(t)}{2\pi}r_{p}\ddot{r}_{p}\!=2\!\int\limits_{0}^{r_{p}}\!\left(\!P\!+\!\frac{B_{z}^{2}}{2\mu_{0}}\!\right)\!rdr.\end{align}
Using the uniform pressure profile in Eq.~(\ref{G:Pl-r}) to evaluate the integral,
\begin{align}\label{G:adiabatic-rp}&\frac{m(t)}{2\pi}r_{p}\ddot{r}_{p}\!\simeq\!-\frac{\mu_{0}I^{2}}{8\pi^{2}}\!+\!P_{l}(t)r_{p}^{2}\!+\!2\!\int\limits_{0}^{r_{p}}\!\frac{B_{z}^{2}}{2\mu_{0}}rdr,\end{align}
which is precisely Eq.~(\ref{Eq:Adiabatic-rp}).
\subsection*{II. Conservation of Energy}
As shown in Appendices~\hyperref[B:IV]{B.IV} and~\hyperref[F:II]{F.II}, the conservation of energy criterion from Eq.~(\ref{F:energy}) still holds with $r_{s}\equiv 0$,
\begin{align}\label{G:energy}&\int\limits_{0}^{r_{p}}\left(\!\frac{2P(r,t)}{\gamma\!-\!1}\!+\!\frac{B_{zl}^{2}(r,t)}{2\mu_{0}}\!\right)rdr\simeq\!\!\int\limits_{r_{p}}^{r_{0}}\!\frac{B_{\theta}^{2}(r_{p}^{+},t)}{2\mu_{0}}r_{p}dr_{p},\end{align}
which can be rearranged as in Eq.~(\ref{F:energy-soln}) to yield,
\begin{align}\label{G:energy-soln}&2\!\int\limits_{0}^{r_{p}}\!\!\left(\!P(r,t)\!+\!\frac{B_{zl}^{2}(r,t)}{2\mu_{0}}\!\right)\!rdr \!=\! \frac{\mu_{0}(\gamma\!-\!1)}{8\pi^{2}}(\Delta(t)\!+\!2\zeta(t))\!\!\nonumber\\
&+\!(3\!-\!\gamma)\!\!\int\limits_{0}^{r_{p}}\!\frac{B_{zl}^{2}(r,t)}{2\mu_{0}}rdr\!+\!\mathcal{O}\!\left(\!r_{0}^{2}P_{0}(r),r_{0}^{2}\frac{B_{z0}^{2}}{2\mu_{0}}\right).\!\end{align}
Using Eq.~(\ref{G:energy-soln}) to evaluate the integral in Eq.~(\ref{G:pre-ddot-rp}) yields the alternative piston ODE,
\begin{align}\label{G:angus-rp}
&\frac{m(t)}{2\pi}r_{p}\ddot{r}_{p}\simeq -\frac{\mu_{0}I^{2}}{8\pi^{2}}\!+\!\frac{\mu_{0}(\gamma\!-\!1)}{8\pi^{2}}(\Delta(t)\!+\!2\zeta(t))\nonumber\\
&+\!(3\!-\!\gamma)\!\!\int\limits_{0}^{r_{p}}\!\frac{B_{zl}^{2}(r,t)}{2\mu_{0}}rdr,\end{align}
where $\Delta(t)$ and $\zeta(t)$ are defined in Eqs.~(\ref{F:delta},\ref{F:zeta}).
The conservation of energy criterion in Eq.~(\ref{G:energy}) can also be used to eliminate the explicit axial field contributions entirely from Eq.~(\ref{G:pre-ddot-rp}) to yield,
\begin{align}\label{G:pre-novel}&\frac{m(t)}{2\pi}r_{p}\ddot{r}_{p}\simeq-\frac{\mu_{0}I^{2}}{8\pi^{2}}+\frac{\mu_{0}}{4\pi^{2}}(\Delta(t)\!+\!2\zeta(t))+\!\frac{\gamma\!-\!3}{\gamma\!-\!1}P_{l}(t)r_{p}^{2}.\end{align}
Once again, Eq.~(\ref{G:angus-rp}) does not explicitly assume a uniform pressure profile and Eq.~(\ref{G:pre-novel}) does not require tracking of the individual fluid elements. They perform similarly to Eq.~(\ref{G:adiabatic-rp}) for realistic scenarios and may be useful alternatives from an implementation standpoint. Nevertheless, Eq.~(\ref{G:adiabatic-rp}) is the simplest, most easily interpretable, and most compatible with Eqs.~(\ref{Eq:Potter},\ref{Eq:simple-rp}) so it is generally preferable as stated in Eq.~(\ref{Eq:Adiabatic-rp}).
\end{document}